%% file: arxiv.tex
\documentclass[11pt]{article}

\pdfoutput=1

\usepackage[utf8x]{inputenc}
\usepackage{amsfonts,amsthm}
\usepackage{amsmath}
\usepackage{complexity}
\usepackage{fullpage}
\usepackage{color}
\usepackage{graphicx}
\usepackage[boxed,noline,longend]{algorithm2e}
\usepackage{bbm}
\usepackage{authblk}
\title{On the Configuration LP for Maximum Budgeted Allocation \thanks{This research was partially supported by ERC
    Advanced investigator grant 226203 and ERC Starting Grant 335288-OptApprox.}}

\author[1]{Christos Kalaitzis}
\author[1]{Aleksander M{\c a}dry}
\author[1]{Alantha Newman}
\author[2]{Luk\'{a}\v{s} Pol\'{a}\v{c}ek}
\author[1]{Ola Svensson}
\affil[1]{EPFL, Switzerland, \texttt{firstname.lastname@epfl.ch}}
\affil[2]{KTH Royal Institute of Technology, Sweden, \texttt{polacek@csc.kth.se}}

\newtheorem{lemma}{Lemma}
\newtheorem{theorem}{Theorem}
\newtheorem{definition}{Definition}

\newtheorem{corollary}{Corollary}

\newtheorem*{neg-corr-restate}{Theorem \ref{thm:neg-corr}}

\newcommand{\hide}[1]{}
\newcommand{\instance}{\ensuremath{\mathcal{I}}}
\newcommand{\instanceJ}{\ensuremath{\mathcal{J}}}
\newcommand{\players}{\ensuremath{\mathcal{A}}}
\newcommand{\items}{\ensuremath{\mathcal{Q}}}
\newcommand{\conf}{\ensuremath{\mathcal{C}}}
\newcommand{\bigitems}{\ensuremath{\mathcal{B}}}
\newcommand{\smallitems}{\ensuremath{\mathcal{S}}}
\newcommand{\Val}{\ensuremath{\mathsf{Val}}}
\newcommand{\Alg}{\ensuremath{\mathsf{Alg}}}
\newcommand{\Avg}{\ensuremath{\mathsf{Avg}}}
\newcommand{\EX}{\ensuremath{\mathbb{E}}}

\newcounter{restatecounter:thm}
\newtheorem{restatement:thm}[restatecounter:thm]{Theorem}

\newtheorem{commentcontainer}{Comment}

\ifthenelse{\isundefined{\DONOTINSERTCOMMENTS}}
{ 
  \newcommand{\comment}[1]%
  {\begin{commentcontainer} \textcolor{blue}{#1} \end{commentcontainer}}
}
{ 
  \newcommand{\comment}[1]{}
}

\graphicspath{{figures/}}

\SetKwFor{ForEach}{foreach}{do}{end}
\SetKwFor{If}{if}{then}{end}

\begin{document}
\maketitle

\begin{abstract}
We study the Maximum Budgeted Allocation problem , i.e.,\, the problem of selling a set of
$m$ indivisible goods to $n$ players, each with a separate budget, such that we maximize the
collected revenue. Since the natural assignment LP is known to have an integrality
gap of 
$\frac{3}{4}$, which
matches the best known approximation algorithms, our main focus is to improve our understanding of the stronger configuration LP
relaxation. In this direction, we prove that the integrality gap of the
configuration LP is strictly better than $\frac{3}{4}$, and provide corresponding polynomial time
roundings, in the following restrictions of the problem: (i) the Restricted Budgeted Allocation
problem, in which all the players have the same budget and every item has the same value for any
player it can be sold to, and (ii) the graph MBA problem, in which an item can be assigned to at
most 2 players. Finally, we improve the best known upper bound on the integrality gap for the
general case from $\frac{5}{6}$ to $2\sqrt{2}-2\approx 0.828$ and also prove hardness of
approximation results for both cases.
\end{abstract}

\input introduction

\input preliminaries

\input old-algorithm

\input restricted

\input neg-corr

\input confround

\input gap

\input hardness

\input future

\bibliography{arxiv}

\end{document}

%% file: introduction.tex
\section{Introduction}
Suppose there are multiple players, each with a budget, who want to pay to gain access to some
advertisement resources. On the other hand, the owner of these resources wants to allocate them so
as to maximize his total revenue, i.e.,\ he wishes to maximize the total amount of money the players
pay.  No player can pay more than his budget so the task of the owner is to find an assignment of
resources to players that maximizes the total payment where each player pays the minimum of his
budget and his valuation of the items assigned to him.


The above problem is called Maximum Budgeted Allocation (MBA), and it arises
often in the context of advertisement allocation systems. Formally, a
problem instance $\instance$ can be defined as follows:
there is a set of players $\players$ and a set of items $\items$. Each
player $i$ has a budget $B_i$ and each item $j$ has a price $p_{ij} \le B_i$
for player $i$ (the assumption that $p_{ij} \le B_i$ is without loss of generality, because no
player can spend more money than his budget). Our objective is to find disjoint sets
$S_i\subseteq\items$ for each player $i$, i.e., an indivisible assignment of items to players,  such that we maximize
$$\sum_{i\in\players}\min\left \{\sum_{j\in S_i}p_{ij},B_i\right\}.$$

In this paper, we are interested in designing good algorithms for the MBA problem and we shall focus
on understanding the power of a strong convex relaxation called the configuration LP. The general goal
is to obtain a better understanding of basic allocation problems that have a wide range of
applications. In particular, the study of configuration LP is motivated by the belief that a deeper
understanding of this type of relaxation can lead to better algorithms for many
allocation problems, including MBA, the Generalized Assignment problem, Unrelated Machine
Scheduling, and Max-Min Fair Allocation.

As the Maximum Budgeted Allocation problem is known to be
NP-hard ~\cite{DBLP:journals/jacm/GandhiKPS06,DBLP:journals/geb/LehmannLN06}, we turn our
attention to approximation algorithms. Recall that an $r$-approximation algorithm is an efficient
(polynomial time) algorithm that is guaranteed to return a solution within a factor $r$ of the
optimal value. The factor $r$ is referred to as the approximation ratio or guarantee.

Garg, Kumar and Pandit~\cite{garg2001approximation}
obtained the first
approximation algorithm for MBA with a guarantee of
$\frac{2}{1+\sqrt{5}}$. This was later improved to $1-\frac{1}{e}$ by
Andelman and Mansour \cite{DBLP:conf/swat/AndelmanM04}, who also
showed that an approximation guarantee of $0.717$ can be obtained in
the case when all the budgets are equal.  Subsequently, Azar,
Birnbaum, Karlin, and Mathieu~\cite{DBLP:conf/icalp/AzarBKMN08} gave a
$\frac{2}{3}$-approximation algorithm, which
Srinivasan~\cite{DBLP:conf/approx/Srinivasan08} extended  to give the best-known approximation
guarantee of $\frac{3}{4}$. Concurrently, the same approximation
guarantee was achieved by Chakrabarty and
Goel~\cite{DBLP:journals/siamcomp/ChakrabartyG10}, who also proved that
it is NP-hard to achieve an approximation ratio better than
$\frac{15}{16}$.

It is interesting to note that the progress on MBA has several points
in common with other basic allocation problems. First, it is observed
that when the prices are relatively small compared to the budgets,
then the problem becomes substantially easier (e.g.
\cite{DBLP:journals/siamcomp/ChakrabartyG10,DBLP:conf/approx/Srinivasan08}),
analogous to how Unrelated Machine Scheduling becomes
easier when the largest processing time is small compared to the
optimal makespan. Second, the above mentioned $3/4$-approximation
algorithms give a tight analysis of a standard LP relaxation, called
assignment LP, which has been a successful tool for allocation problems
ever since the breakthrough work by Lenstra, Shmoys, and
Tardos~\cite{DBLP:journals/mp/LenstraST90}. Indeed, we now have a complete understanding of the strength of the
assignment LP for all above mentioned allocation problems.  The
strength of a relaxation is measured by its integrality gap, which is the
maximum ratio between the solution quality of the exact integer
programming formulation and of its relaxation.

A natural approach for obtaining better (approximation) algorithms for allocation problems are
stronger relaxations than the assignment LP. Similarly to other allocation problems, there is a
strong belief that a strong convex relaxation called a configuration LP gives strong guarantees for
the MBA problem. Even though we only know that the integrality gap is no better than
$\frac{5}{6}$~\cite{DBLP:journals/siamcomp/ChakrabartyG10}, our
current techniques fail to prove that the configuration LP gives even
marginally better guarantees for MBA than the assignment LP. The goal
of this paper is to increase our understanding of the configuration LP
and to shed light on its strength.

\begin{figure}[t]\label{fig:assignmentgap}
\begin{center}
 \includegraphics[width=0.68\textwidth]{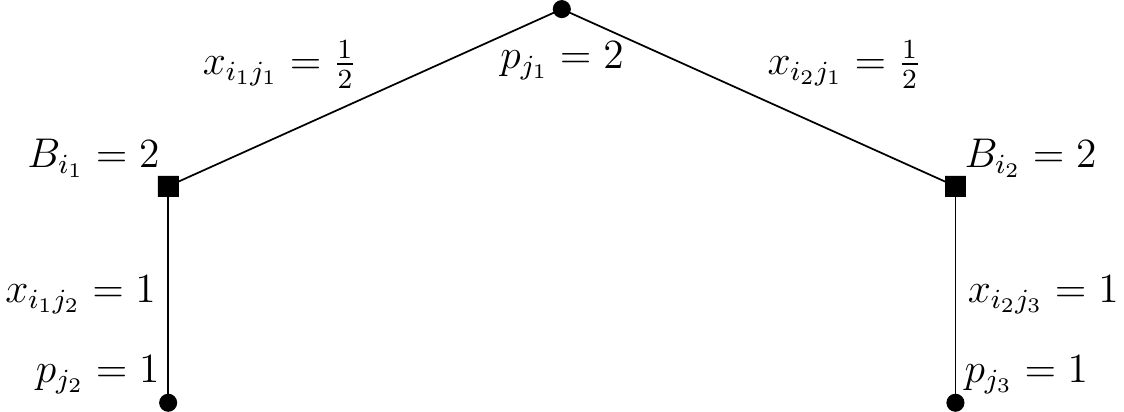}
 \caption{The solution $x$ has value of 4. Any integral solution has a value of at most 3, since one
   player will be assigned only one item of value 1.}
\vspace{-21pt}
\end{center}
\end{figure}

\paragraph{Our contributions.} To analyze the strength of the configuration LP compared to the
assignment LP, it is instructive to consider the tight integrality gap
instance for the assignment LP
from~\cite{DBLP:journals/siamcomp/ChakrabartyG10} depicted in
Figure~\ref{fig:assignmentgap}. This instance
satisfies several structural properties: (i) at most two players have
a positive price of an item, (ii) every player has the same budget
(also known as \emph{uniform budgets}), (iii) the price of an item $j$ for a
player is either $p_j$ or $0$, i.e., $p_{ij} \in \{0,p_j\}$.

Motivated by these observations and previous work on allocation
problems, we shall mainly concentrate on two special cases of MBA. The
first case is obtained by enforcing (i) in which at most two players
have a positive price of an item. We call it \emph{graph MBA}, as an
instance can naturally be represented by a graph where items are
edges, players are vertices and assigning an item corresponds to
orienting an edge. The same restriction, where it is often called
Graph Balancing, has led to several nice results for Unrelated Machine
Scheduling~\cite{ebenlendr2008graph} 
and Max-Min Fair Allocation~\cite{verschae2011configuration}.

The second case is obtained by enforcing (ii) and (iii). That is, each
item $j$ has a non-zero price, denoted by $p_j$, for a subset of
players, and the players have uniform budgets. We call this case
\emph{restricted MBA} or the Restricted Budgeted Allocation Problem as
it closely resembles the Restricted Assignment Problem that has been
a popular special case of both Unrelated Machine
Scheduling~\cite{DBLP:journals/siamcomp/Svensson12} and Max-Min Fair
Allocation~\cite{feige2008allocations,asadpour2008santa,DBLP:conf/stoc/BansalS06}. It
is understood that these two structural properties produce natural
restrictions whose study helps increase our understanding of the
general problem \cite{DBLP:journals/siamcomp/ChakrabartyG10,DBLP:conf/approx/Srinivasan08},
and specifically, instances displaying property (ii) have been studied
in \cite{DBLP:conf/swat/AndelmanM04}.


Our main result proves that the configuration LP is indeed stronger than the assignment LP for the
considered problems.
\begin{theorem}
Restricted Budgeted Allocation and graph MBA have  $(3/4+c)$-approximation algorithms that also
bound the integrality gap of the configuration LP, for some constant $c>0$.
\end{theorem}

The result for graph MBA is inspired by the work by Feige and Vondrak
~\cite{DBLP:conf/focs/FeigeV06} on the generalized
assignment problem and is presented in Section~\ref{sec:graphMBA}. The main idea is to first recover a  $3/4$-fraction of
the configuration LP solution by randomly (according to the LP solution) assigning items to the
players. The improvement over $3/4$ is then obtained by further assigning some of the items that
were left unassigned by the random assignment to players whose budgets were not already exceeded.
The difficulty in the above approach lies in analyzing the contribution of the items assigned in the
second step over the random assignment in the first step (Lemma~\ref{lem:half-integral-items}).

For restricted MBA, we need a different approach. Indeed,  randomly assigning items according to the
configuration LP
only recovers a $(1-1/e)$-fraction of the LP value when an item can be assigned to any number of
players. Current techniques only gain an additional small $\epsilon$-fraction
by assigning unassigned items in the second step. This would lead to an approximation guarantee of $(1-1/e +
\epsilon)$ (matching the result in~\cite{DBLP:conf/focs/FeigeV06} for the Generalized Assignment
Problem) which is strictly less than the best known approximation guarantee of $3/4$ for MBA. We
therefore take a different approach. We first observe that an existing algorithm, described in
Section~\ref{sec:oldalgo}, already gives a better
guarantee than $3/4$ for configuration LP solutions that are not well-structured
(see Definition~\ref{def:wellstructured}). Informally, an LP solution is well-structured if half the budgets of
most players are assigned to expensive items, which are defined as
those items whose price is very close to the budget.  For the rounding of
well-structured solutions in Section~\ref{sec:well-struct}, the main new idea is to first assign expensive/big
items (of value close to the budgets) using random bipartite matching and then assign the cheap/small items
in the space left after the assignment of expensive items.  For this
to work, it is not sufficient to assign the big items in any
way that preserves the marginals from the LP
relaxation. Instead, a key observation is that we can assign big
items so that the probability that two players $i,i'$ are both
assigned big items is at most the probability that $i$ is assigned a
big item times the probability that $i'$ is assigned a big item (i.e.,
the events are negatively correlated).  Using this we can show that we can assign many of the small
items even after assigning the big items leading to the improved guarantee. 
We believe that this is an interesting use of bipartite
matchings for allocation problems as we are using the fact that the events that vertices are matched
can be made negatively correlated. Note that this is in contrast to the events that edges are part
of a random matching which are not necessarily negatively correlated.

Finally, we complement our positive results by hardness results and integrality gaps.  For
restricted MBA, we prove hardness of approximation that matches the strongest results known for the
general case. Specifically, we prove in Section~\ref{sec:hardness} that it is NP-hard to
approximate restricted MBA within a factor $15/16$. This shows that some of the hardest known
instances for the general problem are the ones we study. We also improve the $5/6$ integrality gap
of the configuration LP for the general case: we prove that it is not better than $2(\sqrt{2}-1)$ in
Section~\ref{sec:gap}.

%% file: preliminaries.tex
\section{Preliminaries}

\paragraph{Assignment LP.} The assignment LP for the MBA problem has a fractional ``indicator'' variable
$x_{ij}$ for each player $i \in \players$ and each item $j\in \items$
that indicates whether item $j$ is assigned to player $i$. Recall that
the profit received from a player $i$ is the minimum of his budget
$B_i$ and the total value $\sum_{j\in \items} x_{ij} p_{ij}$ of the
items assigned to $i$. In order to avoid taking the minimum for each
player, we impose that each player $i$ is fractionally assigned items
of total value at most his budget $B_i$. Note that this is not a valid
constraint for an integral solution but it does not change the value
of a fractional solution: we can always fractionally decrease the
assignment of an item without changing the objective value if the
total value of the fractional assignment exceeds the budget. To
further simplify the relaxation, we enforce that all items are fully
assigned by adding a dummy player $\ell$ such that $p_{\ell j} = 0$
for all $j\in \items$ and $B_\ell =0$. The assignment LP for MBA is
defined as follows:

\[
\begin{array}{rrll}
\max &\quad \sum_{i\in \players}\sum_{j\in \items} x_{ij}p_{ij}&\\
\text{subject to} & \sum_{j\in \items} x_{ij}p_{ij}&\leq B_i &\quad \forall i\in \players \\
& \sum_{i\in \players} x_{ij}&=1 & \quad \forall j\in \items\\
& 0 \leq x_{ij} &\leq 1 & \quad \forall i\in \players,\forall j\in \items
\end{array}
\]

As discussed in the introduction, it is known that the integrality gap
of the assignment LP is exactly $\frac{3}{4}$; therefore, in order to
achieve a better approximation, we employ a stronger relaxation called
the configuration LP.

\paragraph{Configuration LP.}  The intuition behind the configuration
LP comes from observing that, in an integral solution, the players are
assigned disjoint sets, or configurations, of items. The configuration
LP will model this by having a fractional ``indicator'' variable
$y_{i\conf}$ for each player $i$ and configuration $\conf \subseteq
\items$, which indicates whether or not $\conf$ is the set of items assigned to
player $i$ in the solution. The constraints of the configuration LP require
that each player is assigned at most one configuration and each
item is assigned to at most one player. If we let $w_i(\conf)=
\min \left\{\sum_{j\in \conf} p_{ij}, B_i\right\}$ denote the total
value of the set $\conf$ of items when assigned to player $i$, the
configuration LP can be formulated as follows:
\[
\begin{array}{rrl}
  \max &\quad \sum_{i\in \players}\sum_{\conf\subseteq \items} w_i(\conf) y_{i\conf}&\\
\text{subject to} & \sum_{\conf\subseteq \items} y_{i\conf}&\leq 1 \quad \forall i\in \players \\
& \sum_{i\in \players, \conf\subseteq \items: j\in \conf} y_{i\conf}&\leq 1 \quad \forall j\in
\items\\
& y_{i\conf}&\geq 0 \quad \forall i\in \players,\forall \conf\subseteq \items
\end{array}
\]
We remark that even though the relaxation has exponentially many variables, it can be solved
approximately in a fairly standard way by
designing an efficient separation oracle for  the dual which has polynomially many variables. We
refer the reader to 
\cite{DBLP:journals/siamcomp/ChakrabartyG10} for more details.

The configuration LP is stronger than the assignment LP  as
it enforces a stricter structure on the fractional solution. Indeed,
every solution to the configuration LP can be transformed into a
solution of the assignment LP of at least the same value (see e.g.
Lemma~\ref{lem:projection}). However, the converse is not true; one example is shown in
Figure~\ref{fig:assignmentgap}. More convincingly, our results show that
the configuration LP has a strictly better integrality gap than the assignment LP for large natural
classes of the MBA problem.

For a solution $y$ to the configuration LP, we let $\Val_i(y) = \sum_\conf w_i (\conf) y_{i
\conf}$ be the value of the fractional assignment to player $i$ and let $\Val(y)=\sum_i \Val_i(y)$. Note
that $\Val(y)$ is equal to the objective value of the solution $y$.
Abusing the notation a little, we also define $\Val_i(x) = \sum_j x_{ij}p_{ij}$ for a solution $x$
to the assignment LP. We might also use $\Val^\instance (y)$ and $\Val_i^\instance (y)$ to make it
clear that we are considering instance $\instance$.

\paragraph{Random bipartite matching.} As alluded to in the introduction, one of the key ideas of
our algorithm for the restricted case is to first assign expensive/big
items (of value close to the budgets) by picking a random bipartite matching so that the events that
vertices are matched are negatively correlated.
The following theorem uses the
techniques developed by Gandhi, Khuller, Parthasarathy and Srinivasan in
their work on selecting random bipartite matchings with particular
properties~\cite{DBLP:journals/jacm/GandhiKPS06}.  For completeness, its
proof is included in Section~\ref{sec:assbigitem}.
\begin{theorem}\label{thm:neg-corr}
Consider a bipartite graph $G=((A,B),E)$ and an assignent $(x_e)_{e\in
  E}$ to edges so that the fractional degree $\sum_{u: uv\in E}
x_{uv}$ of each vertex $v$ is at most $1$. Then there is an efficient,
randomized algorithm that generates a (random) matching satisfying:
\begin{description}

\item[(P1):]{{{\bf Marginal Distribution.}}  For every vertex $v \in A\cup B$, $\Pr[v \mbox{ \emph{is
      matched}}] = \sum_{u: uv \in E} x_{uv}$.  }

\item[(P2):]{{\bf{ Negative Correlation.}} For any $S \subseteq A$,
$
\Pr[\bigwedge_{v \in S} (v\mbox{\emph{ is matched}})]  \leq  \prod_{v \in S} \Pr[v \mbox{ \emph{is matched}}].
$
}
\end{description}
\end{theorem}
One should note that the events \{edge $e$ is in the matching\} and
\{edge $e'$ is in the matching\} are not necessarily negatively
correlated (if we preserve the marginals). A crucial ingredient for
our algorithm is therefore the idea that we can concentrate on the
event that a player has been assigned a big item without regard to 
the specific item assigned.

%% file: old-algorithm.tex
\section{General 3/4-approximation algorithm}
\label{sec:oldalgo}

In this section we introduce an algorithm (inspired by \cite{DBLP:journals/mp/ShmoysT93}) to round
assignment LP solutions and we then present an analysis showing that it is a $3/4$ approximation
algorithm. In Section~\ref{sec:restricted} we use this analysis to show that the
algorithm has a better approximation ratio than $3/4$ in some cases.

First, we need the following definition for the algorithm. Let $G = U\cup V$ be a bipartite graph. A
\emph{complete matching for $V$} is a matching that has exactly one edge incident to every vertex in
$V$.

\begin{algorithm}[h!]
 \DontPrintSemicolon
\SetKwInOut{Input}{Input}\SetKwInOut{Output}{Output}
\Input{Solution $x$ to the assignment LP, ordering $o_i$ of the items by prices for player $i$}
\Output{Assignment $x^*$ of items to the players}
\BlankLine
\ForEach{$i\in \players$}{
  \tcp{Create buckets for player $i$, see Figure~\ref{fig:buckets}}
  $c_i \leftarrow \lceil \sum_j x_{i j} \rceil$\;
  Create $c_i$ buckets $(i, 1), \dots, (i, c_i)$\;
  Create $x'_{(i, \cdot)}$ from $x_i$ as in Figure~\ref{fig:buckets}\;
}
$U \leftarrow \{(i, k) \mid 1 \le k \le \lceil \sum_j x_{ij} \rceil \}$\;
$V \leftarrow \items$\;
Express $x'$ as a convex combination of complete matchings for $V$: $x' = \sum_i
\gamma_i m_i$\;
Return matching $m_i$ with probability $\gamma_i$\;
\caption{Bucket algorithm}
\label{alg:buckets}
\end{algorithm}

\begin{figure}[h]
  \begin{center}
    \includegraphics[scale=0.7]{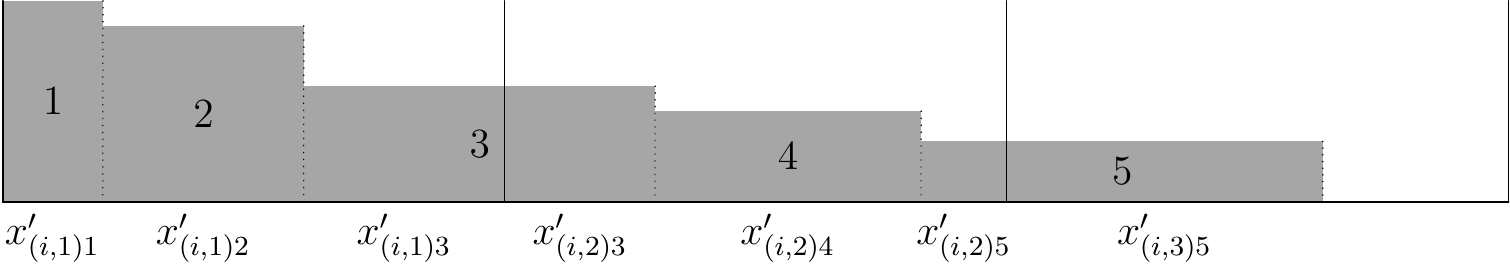}
  \end{center}
  \vspace{-10pt}
  \caption{Illustration of bucket creation by Algorithm~\ref{alg:buckets} for
  player $i$. Buckets are marked by solid lines. The value $x_{i3}$ is split into $x'_{(i,1)3}$ and
  $x'_{(i,2)3}$ and $x_{i5}$ is split into $x'_{(i,2)5}$ and $x'_{(i,3)5}$. For the other items we
  have $x'_{(i, 1)1} = x_{i1}, x'_{(i, 1)2} = x_{i2}, x'_{(i, 2)4} = x_{i4}$. Items are ordered in
  non-decreasing order by their prices.}

  \label{fig:buckets}
  \vspace{-10pt}
\end{figure}

Algorithm~\ref{alg:buckets} first partitions $x$ into buckets creating a new assignment $x'$, such
that the sum of $x'$ in each bucket is exactly 1 except possibly the last bucket of each player.
Some items are split into two buckets. The process for one player is illustrated in
Figure~\ref{fig:buckets}.

From the previous discussion, for every bucket $b$ we have $\sum_j x'_{bj} \le 1$. Also, $\sum_{b\in U}
x'_{bj} = 1$ for every item $j$, which is implied by $\sum_{i\in\players} x_{ij} = 1$ for all
$j\in\items$. Hence $x'$ is inside the complete matching polytope for $V = \items$. Using an
algorithmic version of Carathéodory's theorem (see e.g. Theorem 6.5.11 in
\cite{GroetschelLovaszSchrijver1993}) we can in
polynomial time decompose $x'$ into a convex combination of polynomially many complete matchings for
$V$.

In the algorithm we use an ordering $o_i$ for player $i$ such that $p_{io_{ij}} \ge
p_{io_{i,j+1}}$, i.e.\ it is the descending order of items by their prices for player $i$. In
particular this implies that the algorithm does not use the prices, only the order of items.
Also note that Algorithm~\ref{alg:buckets} could be made deterministic. Instead of picking a random
matching we can pick the best one.

Let $\Alg_i(x)$ be the expected price that player $i$ pays. We know that $\Alg_i(x) \le
\Val_i(x)$, because the probability of assigning $j$ to $i$ is $x_{ij}$, but we
don't have equality in the expression, because some matchings might
assign a price that is over the
budget for some players. In the following lemma we bound $\Alg_i(x)$ from below. 

\begin{lemma}
  Let $x$ be a solution to the assignment LP, $i\in \players$ and let $\alpha \ge 1$ be such that
  $\Val_i(x) = B_i / \alpha$. Let $a_1$ be the average price of items in the first bucket $b$ of player
  $i$, i.e.\ $a_1 = \sum_j x'_{bj} p_{bj}$. Let $r_1$ be the average price
  of items in $b$ that have price more than $\alpha a_1$. Then
  $$\Alg_i(x) \ge \Val_i(x) \left(1 - \frac{r_1}{4 B_i}\right).$$
  In particular, since $r_i \le B_i$, Algorithm~\ref{alg:buckets} gives a 3/4-approximation.
  \label{lem:buckets}
\end{lemma}

\begin{proof}
  The expected value of the solution $x^*$ for player $i$ is $\EX[\Val_i (x^*)] = \Val_i (x)$,
  because the probability of assigning item $j$ to player $i$ is $x_{ij}$. The problem with the
  assignment $x^*$ is that some players might go over the budget, so we cannot make use of the full value
  $\Val_i (x^*)$. We now prove that we only lose $\frac{r_1}{4 B_i}$-fraction of $\Val_i(x)$ by
  going over the budget of player $i$.

  Note that $\alpha \ge 1$, since a solution to assignment LP never goes over the budget $B_i$. Let
  $a_b$ be the average price in bucket $b$. The matching picks at most one item from each bucket and
  suppose from each bucket $b$ it picks item of price more than $\alpha \cdot a_b$. Then since
  $\alpha \sum a_b =
  B_i$, the player would be assigned more than her budget. However, if we assume that all items
  within $b$-th bucket have price at most $\alpha \cdot a_b$, all maximum matchings assign price at
  most $B_i$ to player $i$. We thus define $p'_{bj}$ to be $\min\{p_{bj}, \alpha \cdot a_b\}$ and we
  get a new instance $\instanceJ$ with prices $p'$ and buckets become the players of this instance.
  From the previous discussion we know no player goes over budget when using the fictional prices
  $p'$, so $\Val_i ^ \instanceJ (x^*) \le B_i$. Thus $\Alg_i(x)$ can be lower-bounded by
  $\EX[\Val_i^\instanceJ (x^*)] = \Val_i^\instanceJ (x')$. We now prove that $\Val_i^\instanceJ (x')
  \ge (1 - \frac{r_1}{4 B_i}) \Val_i (x)$.

  Let $r_b$ be the average price of items in bucket $b$ with price above $\alpha \cdot a_b$ and let
  $q_b$ be the average price of the rest of the items in bucket $b$. Note that $q_b \ge r_{b +
  1}$. We let $x'_b$ be the probability corresponding to
  $r_b$, i.e.\ the sum of all $x'_{bj}$ where $p_{bj} > \alpha \cdot a_b$. Since $p_{ij}$ is changed
  to $\alpha \cdot a_b$ in $p'$ for these items, the
  difference in price for bucket $b$ is $(r_b - \alpha a_b) x'_b$. The loss of value by going from
  $p$ to $p'$ is thus
  $$\Val_i (x) - \Val_i ^ \instanceJ (x') = \sum_b (r_b - \alpha a_b) x'_b.$$

  Since $a_b = (r_b - q_b) x'_b + q_b$, we have $(r_b - \alpha a_b) x'_b = (r_b - \alpha q_b)x'_b -
  (r_b - q_b) \alpha {x'_b}^2 \le (r_b - q_b) (x'_b - \alpha {x'_b}^2)$
  where the last inequality follows from $\alpha \ge 1$. Since $q_b \ge r_{b + 1}$, we get $(r_b -
  \alpha a_b) x'_b \le (r_b - q_b) (x'_b - \alpha {x'_b}^2) \le (r_b - r_{b + 1}) (x'_b - \alpha
  {x'_b}^2)$. It follows that
  \begin{equation*}
    \Val_i (x) - \Val_i ^ \instanceJ (x')
    \le \sum_b (r_b - r_{b + 1}) (x'_b - \alpha {x'_b}^2).
  \end{equation*}

  The maximum of $x'_b - \alpha {x'_b}^2$ is attained for $x'_b = \frac{1}{2 \alpha}$ and we get
  \begin{equation*}
    \Val_i (x) - \Val_i^\instanceJ (x') \le \sum_b \frac{r_b - r_{b + 1}}{4 \alpha} \le
    \frac{r_1}{4 \alpha} 
  \end{equation*}

  Hence $$\Alg_i(x) \ge \Val_i^\instanceJ (x') \ge \Val_i(x) - \frac{r_1}{4\alpha} = \Val_i (x)
  \left(1 - \frac{r_1}{4 B_i}\right)$$
  which concludes the proof.
\end{proof}

%% file: restricted.tex
\section{Restricted Budget Allocation}
\label{sec:restricted}

In this section we consider the MBA problem with uniform budgets where the prices are restricted to be
of the form $p_{ij} \in \{p_j, 0\}$. This is the so called \emph{restricted maximum budgeted
allocation}. Our main result is the following.
\begin{theorem}
\label{thm:restbudget}
There is a $(3/4 + c)$-approximation algorithm for restricted MBA for some constant $c > 0$.
\end{theorem}

Since the budgets are uniform, we can assume that each player has a budget of
$1$ by scaling.  We refer to $p_j$ as the price of item $j$ and it is convenient to distinguish whether items
have big or small prices. We let $\bigitems = \{j: p_j \geq 1-\beta\}$ for some $\beta, 1/3 \ge
\beta>0$ to be determined. $\bigitems$ is the set of items of big price and let $\smallitems$
denote the set of the remaining items (of small price).

A key concept for proving Theorem~\ref{thm:restbudget} is that of well-structured solutions; it
allows us to use different techniques based on the structure of the solution to the configuration LP.
In short, a solution $y$ is $(\epsilon, \delta)$-well-structured, if for at least $(1 -
\epsilon)$-fraction of players roughly half of their configurations contain a big item.
\begin{definition}
\label{def:wellstructured}
A solution $y$ to the configuration LP is $(\epsilon, \delta)$-well-structured if
$$
  \Pr_{i} \left[\sum_{\conf \subseteq \items} |\bigitems \cap \conf|\cdot y_{i, \conf}
  \not \in \left[\frac{1-\delta}{2}, \frac{1+\delta}{2}\right]\right] \leq \epsilon,
$$
 where the probability is taken over a weighted distribution of players such that
player $i$ is chosen with probability $\Val_i(y) / \Val(y)$.
\end{definition}

We want to be able to switch from configuration LP to assignment LP without changing the
well-structuredness of the solution. The following lemma shows that it is indeed possible. 

\begin{lemma}
  \label{lem:projection}
  Let $y$ be a well-structured solution to the configuration LP. Then there exists a solution $x$ to
  the assignment LP with $\Val_i(x) = \Val_i(y)$ such that
  \begin{equation*}
    \sum_{\conf \subseteq \items}|\bigitems \cap \conf|\cdot y_{i, \conf} \in
      \left[\frac{1-\delta}{2}, \frac{1+\delta}{2}\right]
    \Leftrightarrow
    \sum_{j: j \in \bigitems} x_{ij} \in \left[\frac{1-\delta}{2}, \frac{1+\delta}{2}\right]
  \end{equation*}
  for all $i \in \players$. Furthermore, $x$ can be produced from $y$ in polynomial time.
\end{lemma}

\begin{proof}
  Note that we can assume that each configuration in $y$ contains at most 2 big items. If a
  configuration contains more than 2 big items, all but 2 big items can be removed without
  decreasing the objective value, since $2(1 - \beta) \ge 1$ and thus the configuration remains over
  the budget.

  We first modify $y$ to a new solution $y'$ where each player does not have at the same time
  a configuration with 2 big items and a configuration with no big items. Solution $y'$ is then
  projected to a solution $x$ to the assignment LP with desired properties.

  Fix player $i$ and two configurations $\conf, \conf'$ such that $\conf$ contains two big items and
  $\conf'$ contains no big items. We can assume that $y_{i\conf} = y_{i\conf'}$, otherwise we can
  split the bigger fractional value into two and disregard one of them. We want to move the second big
  item from $\conf$ into $\conf'$ without decreasing the objective value. This is done by moving
  back small items from $\conf'$.

  Let us order the items in $\conf'$ by value in decreasing order $j_1, \dots, j_k$. We move a
  big item from $\conf$ to $\conf'$ and then move $j_1, j_2, \dots$ to $\conf$ until it has
  profit at least 1 or we run out of items. Let $F$ and $F'$ be the transformed configurations which
  arise from $\conf, \conf'$ respectively.

  If the value of $\conf'$ is less than $\beta$, then we run out of items and the value of $F$ is at
  least $1-\beta + \sum_{j\in\conf'} p_j$ and the value of $F'$ is at least $1-\beta > \beta$, so the objective
  value improved and $y$ is not an optimal solution, a contradiction. Hence we have that the value
  of $\conf'$ is at least $\beta$ and $F$ has value at least $1$.

  It remains to prove that the value of $F'$ is at least as big as the value of $\conf'$. In the
  above process we move a big item of value at least $1-\beta$ to $F'$ and we now show that we
  move less in the other direction. Suppose towards contradiction that we moved more than $1-\beta
  \ge 2\beta$ to $F$, then we moved at least 2 small items and the last item added must have been of
  value at least $\beta$. But then only one such item is necessary, since $\conf$ already contains a
  big item that has value of at least $1-\beta$.

  Applying this procedure whenever we can, we end up with a modified solution $y'$ to the
  configuration LP in which there is no player at the same time having a configuration containing
  two big items and a configuration with no big items. Also, $y'$ is such that
  \begin{equation*}
    \sum_{\conf \subseteq \items}|\bigitems \cap \conf|\cdot y_{i, \conf} =
    \sum_{\conf \subseteq \items}|\bigitems \cap \conf|\cdot y'_{i, \conf},
  \end{equation*}
  because big items are only moved between configurations, so their contribution to the sums above
  is preserved.

  The pairs of configurations $\conf, \conf'$ can be chosen in such a way that we only create
  polynomially many new configurations in total. To see this, let $T = \{\conf_1, \dots, \conf_k\}$ be
  the configurations in $y$ that have two big items and let $S = \{\conf'_1, \dots, \conf'_\ell\}$ be
  the configurations in $y$ with no big items. We process $T$ one by one. For each $\conf_j \in T$
  we try to move the second big item to the configurations $\conf'_1, \dots, \conf'_\ell$ one by
  one. In the end
  we try at most $k\cdot \ell$ different pairs and each pair creates at most 2 new configurations.
  Since $k\cdot \ell$ is polynomial in the size of the instance, also the number of new
  configurations is polynomial.

  Now we project $y'$ to a solution $x$ to the assignment LP as follows: for every player $i$ and
  configuration $\conf$, consider the items in $\conf$ in non-increasing order according to
  $p_{ij}$. If the total value of $\conf$ is at most $B_i$, the configuration $\conf$ contributes
  $y'_{i\conf}$ to the value $x_{ij}$ for all items $j \in \conf$. Otherwise the total value goes
  over the budget and only the first part of the ordered items that is below the budget gets
  contribution $y'_{i\conf}$ and the rest of the items gets contribution 0, except the one item that
  is only partially below the budget which gets some fraction between $0$ and $y'_{i\conf}$. In
  particular this means that big items get the full contribution $y'_{i\conf}$.

  Let us now formalize the intuition given above.
  Let $j'$ be the minimum index such that $\sum_{1\leq j\leq j'} p_{ij}>B_i$. For all
  $j>j'$, set $z_{ij\conf}=0$, for all $j<j'$ set $z_{ij\conf}=y'_{i\conf}$ and set
  $$z_{ij'\conf}=\frac{B_i-\sum_{1\leq j< j'} p_{ij}}{p_{ij'}} y'_{i\conf}.$$
  Finally, we define
  $$x_{ij} = \sum_{\conf \subseteq \items} z_{ij\conf}.$$

  We have $\Val(x) = \Val(y') = \Val(y)$, since $\sum_{j} z_{ij\conf}p_{ij} = w_i(\conf)
  y_{i\conf}$, i.e.\ the contribution of each configuration is preserved by the projection.

  The projection from $y'$ to $x$ gives full contribution $y'_{i\conf}$ to the largest item in $\conf$.
  If there are two big items in $\conf$, the second big item does not get the full contribution.
  However, this happens only when $\sum_{\conf \subseteq \items}|\bigitems \cap \conf|\cdot y'_{i,
  \conf} > 1$ and in this case all configurations in $y'$ for player $i$ have a big item. Then $1
  \le \sum_{j: j \in \bigitems} x_{ij}$. However, if the total weight of big items in $y'$ for
  player $i$ is less than 1, the same total weight is projected on $x$. We thus have
  \begin{equation*}
    \sum_{j: j \in \bigitems} x_{ij} =
    \sum_{\conf \subseteq \items}|\bigitems \cap \conf|\cdot y_{i, \conf}
  \end{equation*}
  if $\sum_{\conf \subseteq \items}|\bigitems \cap \conf|\cdot y_{i, \conf} \le 1$ and otherwise
  $\sum_{j: j \in \bigitems} x_{ij} > 1$. This concludes the proof.
\end{proof}

In the next subsection in Lemma~\ref{lem:non-wellstruct} we show that Algorithm~\ref{alg:buckets}
actually performs better then $3/4$ if the solution $x$ to the assignment LP is produced from a
non-well-structured solution $y$ as in Lemma~\ref{lem:projection}.
In subsection~\ref{sec:well-struct} in Lemma~\ref{lem:wellstruct} we show a new algorithm for
well-structured solutions that also has an approximation guarantee strictly better than $3/4$.
Finally, Lemma~\ref{lem:non-wellstruct} and Lemma~\ref{lem:wellstruct} immediately imply our main
result of this section, i.e., Theorem~\ref{thm:restbudget}.

\subsection{Better analysis for non-well-structured solutions}
We first show that Algorithm~\ref{alg:buckets} performs well if not all players basically are
fully assigned (fractionally). 

\newcommand{\epsa}{\ensuremath{\varepsilon'}}
\newcommand{\epsb}{\ensuremath{\varepsilon''}}

\begin{lemma}
\label{lem:not-fully-assigned}
  Let $\epsa>0$ be a small constant and consider player $i$ such that $\Val_i (x) \leq
  1-\epsa$. Then
  $\Alg_i (x) \ge \frac{3+\epsa/5}{4} \Val_i(x)$.
\end{lemma}
\begin{proof}
For player $i$, select the largest $Q_i$ such that
\begin{equation*}
Q_i = \sum_{j} x_{ij} q_{ij}, \qquad \mbox{ where } q_{ij} = \min\{p_{ij}, Q_i\}.
\end{equation*}

Note that such a $Q_i$ always exists, since $0$ always satisfies the equation. Let $f_i(z) =
\sum_{j} x_{ij} \min\{p_{ij}, z\}$ and note that $f_i(z)$ is a continuous function. We have
$f_i(1) = \Val_i(x) < 1$ and $f_i(0) = 0$ and we want to find the largest $Q_i$, such that $f_i(Q_i) = Q_i$.

If we substitute prices $p_{ij}$ with $q_{ij} = \min\{p_{ij}, Q_i\}$, the ordering of the items for
a player by the price stays the same. Thus Algorithm~\ref{alg:buckets} produces the same result no
matter which one of the two prices we use. Let $D_i$ denote the difference $\sum_j x_{ij} (p_{ij} -
q_{ij}) = \Val_i(x) - Q_i$. We do case distinction based on the size of~$D_i$.

\begin{description}
  \item[Case $D_i > \frac{\epsa}{5} \Val_i(x)$:]\hfill\\
  By Lemma~\ref{lem:buckets}, if we run Algorithm~\ref{alg:buckets} on $x$ but use values $q_{ij}$
  and budget $Q_i$ in the analysis, we get $\Alg_i(x) \ge \frac{3}{4} Q_i$. In order to improve this
  we use the fact that $q_{ij} < p_{ij}$ only for at most one unit of the largest items. If already
  more than one unit of items is at least $Q_i$, then we have $f_i(Q_i) > Q_i$. Since $f_i(1) < 1$
  and $f_i$ is continuous, $Q_i$ is not the largest solution to $f_i(z) = z$, a contradiction.

  Moreover, since the assignment according to the analysis with respect to $q$ does not violate the
  budget $Q_i$, we can always add back the difference $p_{ij} - q_{ij}$ if $j$ is assigned to $i$
  without violating the budget $B_i$, since $p_{ij} - q_{ij} \le B_i - Q_i$. We are not
  losing a quarter from the difference $D_i$, so we have an advantage of $D_i / 4 > \frac{\epsa}{20}
  \Val_i(x)$. Formally, the expected profit is
  \begin{equation*}
  \frac{3}{4}Q_i + D_i = \frac{3}{4}(Q_i + D_i) + D_i/4 = \frac{3}{4} \Val_i(x) + D_i/4,
  \end{equation*}
  which is at least $\left(\frac{3}{4} + \frac{\epsa}{20}\right) \Val_i (x)$.

  \item[Case $D_i \leq \frac{\epsa}{5} \Val_i(x)$:]\hfill\\
  In this case we apply Lemma~\ref{lem:buckets} with prices $q$ and budget $1$. Now since $Q_i$ is
  bounded away from $1$, $\Alg_i(x)$ is more than $3/4Q_i$. Formally,
  \begin{equation*}
  \Alg_i(x) \ge (1 - Q_i/4) Q_i = (1 - Q_i/4) \Val_i (x) - (1 - Q_i/4)D_i \geq (1 - Q_i/4) \Val_i
  (x) - D_i.
  \end{equation*}
  As $Q_i \leq \Val_i (x) \le 1-\epsa$,
  \begin{equation*}
  \Alg_i(x) \ge \left(1- \frac{1-\epsa}{4}\right) \Val_i (x) - D_i
  \geq \left(1- \frac{1-\epsa/5}{4}\right) \Val_i (x) = \frac{3+\epsa/5}{4} \Val_i (x).
  \end{equation*}
\end{description}

\end{proof}

From the above claim, we can see that the difficult players to round are those that have an almost
full budget. Furthermore, we show in the following lemma that such players must have a very special
fractional assignment in order to be difficult to round. 

\begin{lemma}
  Let $\delta > 0$ be a small constant, $\beta$ be such that $\delta / 4 \le \beta$ and consider a
  player $i$ such that $\Val_i(x)
  \geq 1 - \delta^2 / 8$ and $\sum_{j: j \in \bigitems} x_{ij} \not \in \left[\frac{1-\delta}{2},
  \frac{1+\delta}{2}\right]$. Then
  $\Alg_i(x) \ge \frac{3 + \delta^2/64}{4} \Val_i(x)$.
  \label{lem:non-half}
\end{lemma}

\begin{proof}
If the average in the first bucket is more than $\frac{3 + \delta^2/64}{4} \Val_i(x)$ then we are
done, since assigning a random item from that bucket gives sufficient profit.
If
$r_1 \leq 1-\delta^2 / 16$, Lemma~\ref{lem:buckets} already implies the claim. Therefore assume from
now on that $r_1 \ge
1-\delta^2/16$ and $r_2 \le \frac{3 + \delta^2/16}{4}$, so $r_1 - r_2 \geq 1/8$, since $\delta$ is
small.

Select $\alpha \ge 1$ such that $\Val_i(x) = 1/\alpha$. In the proof of Lemma~\ref{lem:buckets} we
have that the expected decrease in value compared to $\Val_i (x)$ in our rounding is at most
$$
\sum_b (r_b - r_{b+1})(x'_b - \alpha {x'_b}^2).
$$

This can be rewritten as
\begin{align*}
(r_1 - r_2) (x'_1 - \alpha {x'_1}^2) + \sum_{b\geq 2} (r_b - r_{b+1})(x'_b - \alpha {x'_b}^2)
&\leq (r_1 - r_2) (x'_1 - \alpha {x'_1}^2) + (r_2 - r_n) \cdot \frac{1}{4 \alpha}\\
&\leq 1/8 (x'_1 - \alpha {x'_1}^2) + 7/8 \cdot \frac{1}{4 \alpha}.\\
\end{align*}
The last inequality follows from $(x'_1 - \alpha {x'_1}^2) \le \frac{1}{4\alpha}$ and $r_1 - r_2\ge
1/8$.

We now prove that $x'_1$ can not be close to $1/2$. The probability $x'_1$ corresponds to items in the
first bucket that have value at least $\alpha a_1$. Suppose towards contradiction that more than
$\delta^2 / (16 \beta)$-fraction of these items are not big items, so they have value at most $1 - \beta$.
Then $r_1 < 1 - \frac{\delta^2}{16 \beta}\cdot \beta = 1 - \delta^2 / 16$, a contradiction. This means
that $x'_1 = (\sum_{j \in \bigitems} x_{ij}) + \gamma$, where $\gamma \in [0, \delta^2/(16\beta)]$.
By $\beta \ge \delta / 4$, $\gamma \in [0, \delta / 4]$. Since $\sum_{j \in \bigitems} x_{ij}
\not\in \left[\frac{1-\delta}{2}, \frac{1+\delta}{2}\right]$, we have $x'_1 \not\in
\left[\frac{1-\delta/2}{2}, \frac{1+\delta/2}{2}\right]$.

We now use the fact that $x'_i$ is bounded away from $1/2$ to prove that the loss in the rounding is
less than $1/4$. For function $z-z^2$ the maximum is attained for $z=1/2$, so $z$ bounded away from
$1/2$ by gives values bounded away from maximum which is 1/4. For function $z-\alpha z^2$ the
maximum is attained very close to $1/2$ provided that $\alpha$ is close to $1$. Again, $z$ bounded
away from $1/2$ gives values bounded away from the maximum. In the rest of the proof we formalize
this intuition.

The maximum for the
function $z(1 - \alpha z)$ is attained for $z = \frac{1}{2 \alpha}$ and we can prove that
$\frac{1}{2 \alpha} \in \left[\frac{1-\delta/2}{2}, \frac{1+\delta/2}{2}\right]$. Since $\alpha \ge 1$,
it only remains to prove that $\frac{1}{2 \alpha} \ge \frac{1 - \delta/2}{2}$. By $1 / \alpha \ge
1 - \delta^2 / 8$,
$$\frac{1}{2 \alpha} \ge \frac{1 - \delta^2 / 8}{2} > \frac{1 - \delta/2}{2}.$$

The function $z - \alpha z^2$ is symmetric around $\frac{1}{2 \alpha}$ and this value is closer to
the beginning of the interval $\left[\frac{1-\delta/2}{2}, \frac{1+\delta/2}{2}\right]$, so the
maximum of $x'_1 - \alpha {x'_1}^2$ is attained when $x'_1 = \frac{1 - \delta/2}{2}$.

We have that
$$x'_1 - \alpha {x'_1}^2 \le x'_1 - {x'_1}^2 = \frac{1 - \delta^2/4}{4} \le \frac{1}{4}{(1 - \delta^2 /
8)}^2.$$

Since $1 - \delta^2 / 8 \le \frac{1}{\alpha}$,
$$x'_1 - \alpha {x'_1}^2 \le \frac{1 - \delta^2 / 8}{4 \alpha}.$$

We can finally bound the decrease in our rounding to be at most
$$
r_1 \left(\frac{1}{8} \cdot \frac{1-\delta^2 / 8}{4 \alpha} + \frac{7}{8} \cdot \frac{1}{4
\alpha}\right) = \frac{r_1 (1 -\delta^2/64)}{4 \alpha}.
$$
The claim follows from the fact that $r_1 \le 1$.
\end{proof}

From Lemma~\ref{lem:not-fully-assigned} and Lemma~\ref{lem:non-half} we have that as soon as a
weighted $\epsilon$-fraction (weight of player $i$ is $\Val_i(y)$) of the players satisfies the
conditions of either lemma, we get a better
approximation guarantee than 3/4. Therefore, when a solution $y$ to the configuration LP is not
$(\epsilon, \delta)$-well-structured, we use Lemma~\ref{lem:projection} to produce a solution $x$ to
the assignment LP for which $\epsilon$-fraction of players satisfies either conditions of
Lemma~\ref{lem:not-fully-assigned} or Lemma~\ref{lem:non-half}. Hence we have the following lemma:

\begin{lemma}
\label{lem:non-wellstruct}
  Given a solution $y$ to the configuration LP which is not $(\epsilon, \delta)$-well-structured and
  $\beta \ge \delta / 4$, we can in polynomial time find a solution with expected value at least
  $\frac{3 + \epsilon \delta^2 / 64}{4}\Val(y)$.
\end{lemma}
\begin{proof}
  Let $x$ be a solution to the assignment LP produced from $y$ as in Lemma~\ref{lem:projection}.
  Then more than weighted $\epsilon$-fraction of players have $\sum_{j: j \in \bigitems} x_{ij} \not\in
  \left[\frac{1-\delta}{2}, \frac{1+\delta}{2}\right]$.

  According to Lemma~\ref{lem:not-fully-assigned} using $\epsilon' = \delta^2 / 8$, we have
  $\Alg_i(x) \geq \frac{3 + \delta^2 / 40}{4} \Val_i(x)$ if $\Val_i(x) \leq 1 - \delta^2 / 8$. By
  Lemma~\ref{lem:non-half}, using $\beta \ge \delta / 4$ implies $\Alg_i(x) \geq \frac{3 + \delta^2 / 64}{4}
  \Val_i(x)$ if $\Val_i(x) \geq 1-\delta^2 / 8$ and $\sum_{j: j \in \bigitems} x_{ij} \not\in
  \left[\frac{1-\delta}{2}, \frac{1+\delta}{2}\right]$. Hence for weighted $\epsilon$-fraction of
  players we get $\Alg_i(x) \geq \frac{3 + \delta^2 / 64}{4} \Val_i(x)$, so the total gain is at
  least $\frac{3 + \epsilon\delta^2 / 64}{4} \Val(y)$.
\end{proof}

\input{well-structured}

%% file: well-structured.tex
\subsection{Algorithm for well-structured solutions}
\label{sec:well-struct}

Here, we devise a novel algorithm that gives an improved approximation
guarantee for $(\epsilon,
\delta)$-well-structured instances when $\epsilon$ and $\delta$ are small constants.

\begin{lemma}
\label{lem:wellstruct}
  Let $1 - \beta$ be the threshold for the big items.  Given a solution $y$ to the configuration LP
  that is $(\epsilon, \delta)$-well-structured, we can in (randomized) polynomial time find a
  solution with expected value at least
$
  (1-\delta)^2(1-\beta-\epsilon) \cdot \frac{25}{32} \Val(y)
$.
\end{lemma}
To prove the above lemma we first give the algorithm and then its
analysis.

\paragraph{Algorithm.} The algorithm constructs a slightly modified
version $y'$ of the optimal solution $y$ to the configuration
LP.  Solution $y'$ is obtained from $y$ in three steps.  First, remove
all players $i$ with $\sum_{\conf \subseteq \items} |\bigitems\cap \conf| y_{i, \conf} \not \in
\left[\frac{(1-\delta)}{2}, \frac{(1+\delta)}{2}\right]$.
As solution $y$ is $(\epsilon, \delta)$-well-structured, this step
decreases the value of the solution by at most~$\epsilon \Val(y)$.

Second, change $y$ as in the proof of Lemma~\ref{lem:projection} by getting rid of configurations
with 2 big items without losing any objective value. Then remove all small items from the
configurations containing big items. After this step, we have the property that big items are alone
in a configuration. We call such configurations big and the remaining ones small. Moreover, we have
decreased the value by at most $\beta \Val(y)$ because each big item has value at least $1-\beta$ and
each configuration has value at most $1$. In the third step, we scale down the fractional assignment
of configurations (if necessary), so as to ensure that $\sum_{\conf: \conf \cap \bigitems =
  \emptyset} y'_{i,\conf} \leq 1/2$ for each player $i \in \players$. As remaining players  are
assigned big configurations with a total fraction at least $(1-\delta)/2$ and therefore small
configurations with a total fraction at most $(1+\delta)/2$, this may decrease the value by a
factor $1/(1+\delta) > 1 - \delta$.

In summary, we have obtained a solution $y'$ for the configuration LP so that each configuration
either contains a single big item or small items; for each remaining player the configurations with
big items constitute a fraction in $\left[\frac{(1-\delta)}{2}, \frac{(1+\delta)}{2}\right]$ and
small configurations constitute a fraction of at most $1/2$. Moreover, $\Val(y') \geq (1- \beta -
\epsilon)(1-\delta)  \Val(y)$.

The algorithm now works by rounding $y'$ in two phases; in the first phase we assign big items and
in the second phase we assign small items.

The first phase works as follows. Let $x'$ be the solution to the assignment LP from
Lemma~\ref{lem:projection} applied on $y'$ and note that $\Val(x') = \Val(y')$. Consider the
bipartite graph where we have a vertex $a_i$ for each player $i\in \players$; a vertex $b_j$ for
each big item $j \in \bigitems$; and an edge of weight $x'_{ij}$ between vertices $a_i$ and $b_j$.
Note that a matching in this graph naturally corresponds to an assignment of big items to players.
We shall find our matching/assignment of big items by using Theorem~\ref{thm:neg-corr}.  Note that by
the property of that theorem we have that (i) each big item $j$ is assigned with probability
$\sum_{i} x'_{ij}$ and (ii) the probability that two players $i$ and $i'$ are assigned big items is
negatively correlated, i.e., it is at most
$\left(\sum_{j\in \bigitems} x'_{ij}\right) \cdot \left(\sum_{j\in \bigitems}
x'_{i'j}\right)$. These two properties are crucial in the analysis of
our algorithm. It is therefore important that we assign the big items according
to a distribution that satisfies the properties of Theorem~\ref{thm:neg-corr}.

After assigning big items, our algorithm proceeds in the second phase to assign the small items as
follows. First, obtain an optimal solution $x^{(2)}$ to  the assignment LP for the small items
together with the players that were
\emph{not} assigned a big item in the first phase; these are the  items that remain and the players
for which the budget is not saturated with value at least $1-\beta$. Then we obtain an integral assignment (of the small
items) of
value at least $\frac{3}{4} \Val(x^{(2)})$ by using Algorithm~\ref{alg:buckets}.

\paragraph{Analysis.} Let $\players_j$ be all the players $i$ for which $x'_{ij} > 0$. Let $x^*$
denote the integral assignment found by the algorithm. Note that the expected value of $x^*$ (over
the randomly chosen assignment of big items) is:
\begin{align*}
\EX[ \Val(x^*)] &= \EX\left[ \sum_{j\in \bigitems}  \sum_{i \in \players_j} x'_{ij} p_j + \frac{3}{4}
  \Val(x^{(2)}) \right] = \sum_{j\in \bigitems}  \sum_{i \in \players_j} x'_{ij} p_j + \frac{3}{4}
\EX[\Val(x^{(2)})].
\end{align*}
We now analyze the second term, i.e., the expected optimal value of the assignment LP where we are
only considering the small items and the set of players $T \subseteq \players$ that
were not assigned big items in the first phase. Then a solution $z$ to the assignment LP can be
obtained by scaling up the fractional assignments of the small items assigned to players in $T$
according to $x'$ by up to a factor of $2$ while maintaining that an item is assigned at most
once. In other words, $z_{ij}=\min\left[1, \sum_{i \in \players_j \cap T} 2x'_{ij} \right]$ and $z$
is a feasible solution to the assignment LP, because we have $\sum_{j \in \smallitems} x'_{ij} p_j
\leq 1/2$.

Thus we have that the expected value of the optimal solution to the assignment LP is by linearity of
expectation is at least
$
\EX_T[\Val(x^{(2)})] \geq \sum_{j \in \smallitems} p_j  \cdot \EX_T \left[ \min \left[ 1, \sum_{i \in
\players_j \cap T} 2x'_{ij} \right]\right]
$.

We continue by analyzing the expected fraction of a small item present in the constructed solution
to the assignment LP. In this lemma we use that the randomly selected matching of big
  items has negative correlation. To see why this is necessary, consider a small item $j\in
  \smallitems$ and suppose that $j$ is assigned to two players $A$ and $B$ both with a fraction
  $1/2$, i.e., $x'_{Aj} = x'_{Bj} = 1/2$. As the instance is well-structured both $A$ and $B$ are
  roughly assigned half a fraction of big items; for simplicity assume it to be exactly $1/2$.
  Note that in this case we have that $ \min \left[ 1, \sum_{i \in \players_j \cap T} 2x'_{ij}
  \right]$ is equal to $1$ if not both $A$ and $B$ are assigned a big item and $0$ otherwise.
  Therefore, on the one hand, if the
  event that $A$ is assigned a big item and the event that $B$ is assigned a big item were perfectly
  correlated then we would have $ \EX_T \left[ \min \left[ 1, \sum_{i \in \players_j \cap T}
      2x'_{ij} \right]\right] =1/2$.  On the other hand, if those events are negatively correlated then $ \EX_T \left[ \min \left[ 1, \sum_{i \in \players_j \cap T}
      2x'_{ij} \right]\right] \geq 3/4$, as in this case the probability that both $A$ and $B$ are
  assigned big items is at most $1/4$. 
\begin{lemma}
\label{lem:min-wellstruct}
For every $j\in \smallitems$, $\EX_T  \left[ \min \left[1, \sum_{i \in \players_j \cap T} 2x'_{ij}
\right]\right] \geq (1-\delta) \frac{3}{4}
  \sum_{i\in \players_j} x'_{ij}$.
\end{lemma}
\begin{proof}
We slightly abuse notation and also denote by $T$ the event that the players in $T\subseteq
\players$ were those that were not assigned big items. Let also $x(T) = \min \left[1, \sum_{k \in
\players_j \cap T} 2x'_{kj} \right]$ for the considered small item $j$. With this notation,
$$
\EX_T  \left[ \min \left[1, \sum_{i \in \players_j \cap T} 2x'_{ij} \right]\right] = \sum_{T
  \subseteq \players} \Pr[T] \cdot x(T).
$$
We shall show that we can lower bound this quantity by assuming that $j$ is only fractionally
assigned to two players. Indeed, suppose that $j$ is fractionally assigned to more than two players.
Then there must exist two players, say $i$ and $i'$, so that $0 <x'_{ij} <
1/2$ and $0 < x'_{i'j}<1/2$; the fractional assignment of a small item to some player never
exceeds $1/2$ by construction of $y'$ and $x'$.
We can write $\sum_{T
  \subseteq \players}\Pr[T] \cdot x(T)$ as
\begin{align}
\label{eq:summy}
\sum_{T  \subseteq \players \setminus \{i,i'\}} ( &\Pr[T] \cdot x(T) + \Pr[T \cup \{i\}] \cdot
  x(T \cup \{i\}) + \Pr[T \cup \{i'\}] \cdot
  x(T \cup \{i'\}) \\
\nonumber & +  \Pr[T \cup \{i,i'\}] \cdot
  x(T \cup \{i,i'\}))
\end{align}
Note that if we shift some amount of fractional assignment from $x'_{ij}$ to $x'_{i'j}$ (or
vice-versa) then $x(T)$ and $x(T \cup \{i,
i'\})$ do not change. We shall now analyze the effect such a shift has on the sums $\sum_{T
  \subseteq \players \setminus \{i,i'\}}   \Pr[T \cup \{i\}] \cdot
  x(T \cup \{i\})$ and  $\sum_{T
  \subseteq \players \setminus \{i,i'\}}   \Pr[T \cup \{i'\}] \cdot
x(T \cup \{i'\})$.
Note that after this shift $x'$ might not be a valid solution to the assignment LP, namely we might
go over the budget for some players. However, our goal is only to prove a lower-bound on
$\EX[\Val(x^{(2)}]$.

For this purpose let $F_i$ denote the probability that the set $T$ is selected such that the value
of
$x(T\cup \{i\})$ is strictly less than $1$, i.e.,
$$
F_i := \sum_{T
  \subseteq \players \setminus \{i,i'\} : x(T \cup \{i\}) < 1}   \Pr[T \cup \{i\}].
$$
Similarly we define $G_i$ for sets where $x(T\cup \{i\})$ is exactly 1, i.e.,
$$
G_i := \sum_{T
  \subseteq \players \setminus \{i,i'\} : x(T \cup \{i\}) = 1}   \Pr[T \cup \{i\}].
$$
The definition of $F_{i'}$ and $G_{i'}$ is analogous. Note that if we, on the one hand, decrease
$x'_{ij}$ by a small $\eta$ and increase $x'_{i'j}$ by $\eta$, this changes~\eqref{eq:summy} by $\eta
(-F_i - G_i + F_{i'})$. On the other hand, if we increase $x'_{ij}$ and decrease $x'_{i'j}$ by $\eta$,
then~\eqref{eq:summy} changes by $\eta (F_i - F_{i'} - G_{i'})$. We know that one of $\eta(F_i -
F_{i'})$ and $\eta(-F_i + F_{i'})$ is non-positive,
so either $\eta (-F_i - G_i + F_{i'})$ or $\eta (F_i - F_{i'} - G_{i'})$ are non-positive as well.

After the small change by $\eta$, $F_i$ increases by $G_i$ or $F_{i'}$ by $G_{i'}$, so further
changes in the same direction will be also non-positive. We can therefore either shift fraction of
$x'_{ij}$ to $x'_{i'j}$ (or
vice versa) without increasing~\eqref{eq:summy} until one of the variables either becomes $0$ or
$1/2$. If it becomes $0$ then we repeat  with one less fractionally assigned small item and if it
becomes
$1/2$ then we repeat by considering two other players where $j$ is fractionally assigned strictly
between $0$ and $1/2$.

By repeating the above process, we may thus assume that $j$ is fractionally assigned to at most
two players say $1$ and $2$ and $x'_{1j}, x'_{2j} \leq 1/2$. We therefore have that \eqref{eq:summy}
is equal to
$$
\Pr[ 1,2 \not \in T] \cdot 0  + \Pr[ 1 \in T, 2 \not \in T]\cdot  2x'_{1j} + \Pr[
1
\not \in T, 2 \in T]\cdot  2x'_{2j} + \Pr[1, 2\in T] \cdot \min[ 1, 2x'_{1j} +2x'_{2j}]
$$

It is clear that the above expression is minimized whenever $ \Pr[1, 2\in T]$ is maximized;
however, since our distribution over the assignments of big items is negatively correlated and it
preserves
the marginals (which are at most $(1+\delta)/2$), it holds that $ \Pr[1, 2\in T]\leq
\frac{(1-\delta)^2}{4}$(since the worst case is that $\Pr[1\in T]=\Pr[2\in T]=\frac{1-\delta}{2}$),
and hence one can see that the above expression is at least
$$
\left(\frac{1-\delta}{2}-\frac{(1-\delta)^2}{4}\right)2x'_{1j}+\left(\frac{1-\delta}{2}-\frac{
\left(1-\delta\right)^2}{4}\right)2x'_{2j
}+\frac{(1-\delta)^2}{4} \min[1, 2x'_{1j} + 2x'_{2j}] =
$$
$$
\frac{1+\delta}{2} \frac{1-\delta}{2} \cdot 2x'_{1j} +  \frac{1+\delta}{2} \frac{1-\delta}{2} \cdot
2x'_{2j} +  \frac{(1-\delta)^2}{4} \min[1, 2x'_{1j} + 2x'_{2j}]
$$
which is at least
$$
\frac{2-2\delta^2}{4} (x'_{1j} + x'_{2j}) + \frac{1-2\delta + \delta^2}{4} (x'_{1j} + x'_{2j}) \geq
(1-\delta) \frac{3}{4}  (x'_{1j} + x'_{2j}).
$$
\end{proof}

Let us now see how it implies Lemma~\ref{lem:wellstruct}. We have that $\EX[ \Val(x^*)]$ is equal to
\begin{align*}
& \sum_{j\in \bigitems}  \sum_{i \in \players_j} x'_{ij} p_j + \frac{3}{4}
  \EX[\Val(x^{(2)})] \geq\  \sum_{j\in \bigitems}  \sum_{i \in \players_j} x'_{ij} p_j +  (1-\delta)
  \left(\frac{3}{4}\right)^2 \sum_{j \in \smallitems} \sum_{i \in \players_j} x'_{ij} p_j.
\end{align*}
As $\sum_{j \in \bigitems} x'_{ij} \geq \frac{1-\delta}{2}$  for every remaining player, we have
$$
\frac{\EX[\Val(x^*)]}{\Val(x')}  \geq (1-\delta) \left(\frac{1}{2} + \frac{1}{2} \frac{9}{16}
\right) = (1-\delta) \frac{25}{32}.
$$
Lemma~\ref{lem:wellstruct} now follows from  that $\Val(x') \geq (1- \beta - \epsilon)(1-\delta)
\Val(y)$. We have proved  Lemmas \ref{lem:non-wellstruct} and \ref{lem:wellstruct}, which in turn imply
Theorem \ref{thm:restbudget} and our analysis is concluded.

%% file: neg-corr.tex
\section{Assigning Big Items}\label{sec:assbigitem}

In this section, we show how to find a matching in a bipartite graph
$G=(A,B,E)$ such that, for any set of vertices $S \subset A$, the
probability that all the vertices of the set are matched is negatively
correlated.  We use the dependent rounding scheme of Gandhi et
al.~\cite{DBLP:journals/jacm/GandhiKPS06}.  Because our precise goal
differs from what they consider, we must slightly modify their analysis.

\begin{definition}
Let $G=((A,B),E)$ be a bipartite graph with edge weights $x:E
\rightarrow [0,1]$.  We say that $G$ is a {\em normal} bipartite graph
if for all $i \in A \cup B, ~\sum_{j:(i,j) \in E} x_{ij} \leq 1$.
\end{definition}
Note that $|A|$ and $|B|$ may not be equal.  Suppose we have a
randomized algorithm that produces a matching $M$.  Let $X_{ij}$
denote the random variable that is 1 if edge $(i,j)$ belongs to $M$
and is 0 otherwise.  Let $X_i$ denote the random variable that is 1 if
vertex $i$ is matched in $M$ and 0 otherwise.  We will show that there
is a randomized algorithm to generate a matching $M$ such that the
following properties hold.
\begin{itemize}

\item[]{{\bf{(P1): Marginal Distribution.}}  For every vertex $i \in
  A\cup B$, $\Pr[X_{i} = 1] = \sum_{j: (i,j) \in E} x_{ij}$.  }

\item[]{{\bf{(P2): Negative Correlation.}} Let $S \subseteq A$.  
\begin{eqnarray*}
\Pr[\bigwedge_{i \in S} (X_i = 1)] & \leq & \prod_{i \in S} \Pr[X_i = 1].
\end{eqnarray*}
}
\end{itemize}

\begin{neg-corr-restate}
If $G=((A,B),E)$ is a {\em normal} bipartite graph, then there is an efficient,
randomized algorithm that generates a matching $M$ satisfying
properties (P1) and (P2).
\end{neg-corr-restate}

\subsection{Algorithm and Proof of Theorem \ref{thm:neg-corr}}

We show that the dependent rounding scheme from
\cite{DBLP:journals/jacm/GandhiKPS06} yields a matching algorithm that
has the desired properties.  We include the algorithm here for
completeness.  First, we give a short description.

The initial step of the algorithm is to remove all cycles from $G$
while preserving the sums at each vertex.  In other words, it first
preprocesses the $y_{ij}$ values (note that initially, $y_{ij} =
x_{ij}$ for all edges in $G$) so that for each vertex $i \in A \cup
B$, the value $\sum_{j:(i,j) \in E} y_{ij}$ is preserved, but the
resulting graph contains no cycles.  This can be done since $G$ is
bipartite and may contain only even cycles, and is implemented in Step
\ref{alg:cycles} of the algorithm.

After Step \ref{alg:cycles}, the graph $G$ has been modified so that
it no longer contains cycles.  However, it may be the case that some
of the $y_{ij}$ values are fractional.  Our goal is to make all of
these values integral, so that the $y_{ij}$ values correspond to a
matching and the sums of the values corresponding to the edges
adjacent a vertex are preserved in expectation.  It is important to
note that these sums may be less than 1.  To obtain a matching from an
acyclic graph, we use the method from
\cite{DBLP:journals/jacm/GandhiKPS06}.  We choose a path (which can be
either even or odd in length) and divide this path into two matchings.
The current $y_{ij}$ edge values restrict how much the edges can be
increased or decreased (we never want any $y_{ij}$ value to exceed 1
or to be negative).  We use these bounds to increase and decrease the
$y_{ij}$ values in this path so as to preserve the expected value of
the sums of the $y_{ij}$ values adjacent to a vertex.

\vspace{5mm}

\begin{figure}[t!]
\noindent
\begin{small}
\fbox{\parbox{16cm}{

{\sc{Matching Algorithm}}

{\it Input:} A normal bipartite graph $G=((A,B),E)$ with edge weights
$x:E \rightarrow (0,1)$.

{\it Output:} A matching $M \subset E$.

\begin{enumerate}

\item Initialize each edge weight with value $y_{ij} := x_{ij}$.

\item If $y_{ij} = 1$, then add edge $ij$ to the matching $M$.  If
  $y_{ij} = 0$, then delete edge $(i,j)$ from $G$.

\item \label{alg:cycles} Remove all cycles: A cycle $C$ can be decomposed into two matchings,
  $M_1$ and $M_2$.  Let $\alpha = \min_{h\ell \in C} x_{h \ell}$ and without
  loss of generality, assume that $(i,j) \in M_1$, where $\alpha = x_{ij}$.  

  \begin{itemize}

  \item Set $y_{ij} := y_{ij} - \alpha$ for all $(i,j) \in M_1$ and
    $y_{ij} := y_{ij} + \alpha$ for all $(i,j) \in M_2$.

    \item If $y_{ij} = 1$, add edge $(i,j)$ to $M$.
      Delete all edges with integral $y_{ij}$ values from $G$.\\
({\em Note that this procedure does not change the fractional degree
  at any vertex.})

  \end{itemize}

\item While $G$ is non-empty: Find a maximal path $P$ in $G$.

\begin{itemize}

\item Divide $P$ into matchings
  $M_1$ and $M_2$.

\item {Choose $\alpha$ and $\beta$ as follows:

  \begin{itemize}

  \item[] $\alpha = \min\{\gamma > 0:  (\exists (i,j) \in M_1: y_{ij}
    + \gamma = 1) \bigvee (\exists (i,j) \in M_2: y_{ij} - \gamma =
    0)\}$;
    
  \item[] $\beta = \min\{\beta > 0:  (\exists (i,j) \in M_1: y_{ij}
    - \gamma = 1) \bigvee (\exists (i,j) \in M_2: y_{ij} + \gamma = 0)\}$.
    
  \end{itemize}
}

\item {With probability $\beta/(\alpha + \beta)$, set: 

  \begin{itemize}

  \item[] $y_{ij} := y_{ij} + \alpha$ for all $(i,j) \in M_1$ and
    $y_{ij} := y_{ij} - \alpha$ for all $(i,j) \in M_2$.

  \end{itemize}}

\item {With probability $\alpha/(\alpha + \beta)$, set:

  \begin{itemize}

  \item[] $y_{ij} := y_{ij} - \beta$ for all $(i,j) \in M_1$ and
    $y_{ij} := y_{ij} + \beta$ for all $(i,j) \in M_2$.

  \end{itemize}}

\item If $y_{ij} = 1$, add edge $(i,j)$ to $M$.  Delete all edges with
  integral $y_{ij}$ values from $G$.

\end{itemize}


\end{enumerate}

}}
\end{small}
\end{figure}

\subsection{Analysis}

We note that the {\sc Matching Algorithm} takes no more than $|E|$
rounds.  This is because at each step, at least one edge becomes
integral and is therefore removed from the edge set.  

\begin{lemma}
Property (P1) holds.
\end{lemma}

\begin{proof}
Suppose the algorithm runs $t$ iterations of Step 4.  Fix edge $(i,j)
\in E$.  Let $Y_{ij,k}$ denote the value of $y_{ij}$ right before
iteration $k$.  We will show that:
\begin{eqnarray}
\forall k \geq 1, ~ \mathbbm{E}[Y_{ij,k+1}] = \mathbbm{E}[Y_{ij,k}].\label{holds}
\end{eqnarray}
Note that $Y_{ij,1}$ denotes the value of $y_{ij}$ after the last execution
of Step 3 and before the first execution of Step 4.
Then we have $\Pr[X_{ij} = 1] = \mathbbm{E}[Y_{ij,t}] =
\mathbbm{E}[Y_{ij,1}]$.  Note that $\sum_{j:(i,j) \in E}
x_{ij} = \sum_{j:(i,j) \in E} Y_{ij,1}$, since in removing cycles, the
degree sums do not change.

Suppose edge $(i,j) \notin P$ where $P$ is the path chosen in round
$k$.  Then Equation \eqref{holds} holds.  If edge $(i,j) \in P$ and in
$M_1$, then
we have:
\begin{eqnarray}
\mathbbm{E}[Y_{ij,k+1}] & = & \frac{\beta}{\alpha+\beta}
(\mathbbm{E}[Y_{ij,k}] + \alpha) +
\frac{\alpha}{\alpha+\beta}(\mathbbm{E}[Y_{ij,k}] - \beta)\\
& = & \mathbbm{E}[Y_{ij,k}].
\end{eqnarray}
If edge $(i,j) \in P$ and in $M_2$, then we have:
\begin{eqnarray}
\mathbbm{E}[Y_{ij,k+1}] & = & \frac{\beta}{\alpha+\beta}
(\mathbbm{E}[Y_{ij,k}] - \alpha) +
\frac{\alpha}{\alpha+\beta}(\mathbbm{E}[Y_{ij,k}] + \beta)\\
& = & \mathbbm{E}[Y_{ij,k}].
\end{eqnarray}
\end{proof}

\begin{lemma}
Property (P2) holds.
\end{lemma}

\begin{proof}
Let $Y_{i,k}$ denote that value of $\sum_{j \in B}
y_{ij}$ before round $k$.  We can show:
\begin{eqnarray}
\forall k \geq 1, \mathbbm{E}\left[\prod_{i \in S} Y_{i,k+1} \right] & \leq &
\mathbbm{E}\left[\prod_{i \in S} Y_{i,k} \right]. \label{never-change}
\end{eqnarray}
Let $t$ be the last round of the algorithm.
If Equation \ref{never-change} holds, then we have the following:
\begin{eqnarray}
\Pr[\bigwedge_{i \in S} (X_i = 1)] & = & \mathbbm{E}\left[\prod_{i \in
    S} Y_{i,t} \right]\\
& \leq & \mathbbm{E}\left[\prod_{i \in S} Y_{i,1} \right]\\
& = & \prod_{i \in S} \sum_{j \in B} x_{ij} \\
& = & \prod_{i \in S} \Pr[X_i = 1].
\end{eqnarray}
Let us now prove Equation \eqref{never-change}.  Consider a maximal
path $P$ in round $k$ and consider the quantity 
\begin{eqnarray}
\mathbbm{E}\left[\prod_{i \in S} Y_{i,k} \right].
\end{eqnarray}
Note that for any node $i \in S$ such that $i$ is an internal
node of $P$, the quantity $Y_{i,k} =
Y_{i,k+1}$.  Let us consider the two endpoints of $P$, which we refer
to as $p_1$ and $p_2$.  If $p_1$ and $p_2$ are both in $S$, then if
the edge adjacent to $p_1$ in $M$ increases, the edge adjacent to
$p_2$ in $M$ will decrease, and vice-versa.
Thus, Equation \eqref{never-change} will be  implied by the
following:
\begin{eqnarray}
\mathbbm{E}\left[Y_{p_1,k+1} \cdot Y_{p_2,k+1}\right] & \leq & 
\mathbbm{E}\left[Y_{p_1,k} \cdot Y_{p_2,k}\right].
\end{eqnarray} 
For some fixed set $S \subseteq A$, $\mathbbm{E}\left[Y_{p_1,k+1} \cdot Y_{p_2,k+1}\right] $ is
equal to
\begin{eqnarray}
&  &
\mathbbm{E}\left[\frac{\beta}{\alpha+\beta}(Y_{p_1,k} - \alpha)(Y_{p_2,k} + \alpha) +
\frac{\alpha}{\alpha+\beta}(Y_{p_1,k} + \beta)(Y_{p_2,k} -
\beta)\right] \\ 
& = & \mathbbm{E}\left[Y_{p_1,k}\cdot Y_{p_2,k} - \frac{\alpha^2\beta}{\alpha+\beta} -
  \frac{\beta^2\alpha}{\alpha+\beta} \right]\\
& \leq & \mathbbm{E}\left[Y_{p_1,k} \cdot Y_{p_2,k} \right].
\end{eqnarray}

If $p_1 \in S$ and $p_2 \notin S$, then Equation
\eqref{never-change} will be implied if the following holds:
\begin{eqnarray}
\mathbbm{E}[Y_{p_1,k+1}] & \leq & \mathbbm{E}[Y_{p_1,k}]. \label{last-ineq1}
\end{eqnarray}
If the edge adjacent to $p_1$ is in $M_1$, then we have:
\begin{eqnarray}
\mathbbm{E}[Y_{p_1,k+1}] & = & \mathbbm{E}[\frac{\beta}{\alpha+ \beta} (Y_{p_1,k}
+ \alpha) + \frac{\alpha}{\alpha+ \beta}(Y_{p_1,k} - \beta)] \\
& = & \mathbbm{E}[Y_{p_1,k}].
\end{eqnarray}
If the edge adjacent to $p_1$ is in $M_2$, then the calculation is analogous.
\end{proof}

%% file: confround.tex
\section{An algorithm for graph MBA}
\label{sec:graphMBA}

In this section we consider graph MBA. Specifically, every player $i\in
\players$ has a (possibly different) budget $B_i$ and every item
$j\in \items$ can be assigned to two players $i,i'$ with a price of
$p_{ij},p_{i'j}$ respectively. Thus this can be viewed as a graph
problem where items are edges, players are vertices and assigning an
item means directing an edge towards a vertex.
 
For this problem, we already know that the integrality gap of the
assignment LP is exactly
$\frac{3}{4}$~\cite{DBLP:conf/swat/AndelmanM04}, and that of the
configuration LP is no better than
$\frac{5}{6}$~\cite{DBLP:journals/siamcomp/ChakrabartyG10}. We prove
that using the configuration LP, we can recover a fraction of the LP
value that is bounded away from $\frac{3}{4}$ by a constant, implying
the following theorem:
 
\begin{theorem}\label{thm:threshold}
There is a polynomial time algorithm which returns a $(\frac{3}{4}+c)$-approximate solution
to the
graph MBA problem, for some constant $c>0$.
\end{theorem}
 
\subsection{Description of the algorithm}
In order to introduce the algorithm, we need to define some notation
first. Let $y$ be a solution to the configuration LP. We abuse
notation and use $x$ to denote a fractional assignment such that for
all $i\in\players$ and $j\in\items$, $
x_{ij}=\sum\limits_{\conf\subseteq \items: j\in
\conf}y_{i\conf}$. Note that we can always maintain that for all
$j\in \items$, $\sum\limits_{i\in \players} x_{ij}=1$, by assigning
item $j$
to some arbitrary configuration if need be, even though that
configuration might already have value which is higher than the
budget of the player to which it is assigned.

In order to argue about the value we recover from the items, for every player $i$ and every
configuration $\conf$, consider the
items in $\conf$ in non-increasing order with respect to their value for player $i$; then
$\conf=\{j_1...j_k\}$, where $k$ is the number of items contained in $\conf$. Consider the minimum
index $\ell$ such that $\sum\limits_{j_t: 1\leq t\leq \ell-1}p_{ij}<B_i$ and $\sum\limits_{j_t:
1\leq t\leq \ell}p_{ij}\geq B_i$; we will call items $\{j_1...j_{\ell-1}\}$ to be below the budget
for configuration $\conf$, items $\{j_{\ell+1}...j_k\}$ to be above the budget and item $j_\ell$ to
be an $f$-fraction below the budget, where
$$f=\frac{B_i-\sum\limits_{j_t: 1\leq t\leq \ell-1}p_{ij_t}}{p_{ij_\ell}}.$$
Consider the contribution of an item to the
value of a configuration,
denoted by $p_{ij\conf}$; we consider that $p_{ij\conf}=p_{ij}$ for items that are
entirely
below the budget, $p_{ij\conf}=0$ for items that are entirely above the budget and
$p_{ij\conf}=fp_{ij}$ for items of which an $f$-fraction is below the budget. Now, let
$\Val_{ij}(y )=\sum\limits_{\conf\subseteq \items:j\in \conf}
y_{i\conf}p_{ij\conf}$ be the contribution of item $j$ to the LP objective value which comes from
its assignment to player $i$, and let $\Val_{j}(y)=\sum\limits_{i\in
\players}\Val_{ij}(y)$ be the contribution of item $j$ to the objective value of the LP. Let
$ \Val(y)=\sum\limits_{j\in \items}\Val_j(y)$ be the value of the
fractional solution $y$ and let $\Val(y^*)=\sum\limits_{j\in \items}w_j(y^*)$ be the value of the
rounded solution $y^*$.

We overload notation and use $\Val(z),\Val_j(z)$ and $\Val_{ij}(z)$,
where $z$ is an assignment (possibly fractional). Since the only fractional assignment we
encounter is directly derived from the configuration LP solution, every assignment we
encounter has a corresponding configuration LP solution and therefore it makes sense to use $\Val$
for fractional assignments and configuration LP solutions interchangeably. Finally, let
$\Avg_{ij}(x)=\frac{\Val_{ij}(x)}{x_{ij}}$ be the
average contribution of item $j$ over all configurations $\conf$ it belongs to in player $i$, such
that $y_{i\conf}>0$; again, we
will overload notation and use both $\Avg_{ij}(z)$ and $\Avg_{ij}(z')$, where $z$ is a fractional
assignment and $z'$ a solution to the configuration LP. Then,
$\Val_j(y)=x_{ij}\Avg_{ij}(y)+x_{i'j}\Avg_{i'j}(y)$.
 
We now present Algorithm \ref{alg:confround}
(notice the similarity
between the secondary assignment used here and the techniques used to tackle GAP in
\cite{DBLP:conf/focs/FeigeV06}). It works in two phases: during the first phase, we choose one
configuration at random for every player, such that player $i$ chooses configuration $\conf$ with
probability $y_{i\conf}$. Certain items might be picked more than once, and these
conflicts are resolved as follows: if item $j$ is picked by both players $i,i'$ we assign it to $i$
with probability $x_{i'j}$, and vice versa. The idea behind this scheme is to penalize players
which are more likely to cause a conflict; the result of this scheme is that we can place a lower
bound not only on the value recovered by every item but also on the value recovered by every
player.We call this phase primary assignment.
 
During the second phase, we want to allocate unassigned items; the way we do this is by assigning
unassigned item $j$ to the player $i$ which has the maximum $\Val_{ij}(x)$. We call this phase
secondary assignment.
 
\begin{algorithm}[h!]
\DontPrintSemicolon
\SetKwInOut{Input}{Input}\SetKwInOut{Output}{Output}
\Input{Solution $y$ to the configuration LP}
\Output{Assignment $x^*$ of items to the players}
\BlankLine
\ForEach{$i\in \players$}{
\tcp{Primary assignment}
Pick a configuration $\conf$ with probability $y_{i\conf}$
}
\ForEach{$j\in \items$}{
\tcp{Conflict resolution}
\If{$j$ is chosen by $i,i'$}{
assign $j$ to $i$ with probability $x_{i'j}$, otherwise assign $j$ to $i'$
}
 
\tcp{Secondary assignment}
\If{$j$ is unassigned}{
assign $j$ to $i\in\players$ which maximizes $\Val_{ij}(x)$
}
}
Return assignment of items $x^*$\;
\caption{Configuration sampling algorithm}
\label{alg:confround}
\end{algorithm}
 
The main idea of the algorithm is that when all the item assignments
are away from 1 and 0, we are guaranteed that some constant fraction
of the budget will be left empty on every player (in expectation), and
thus we can use this fraction to increase the contribution of
unassigned items (since otherwise we could not guarantee that their
contribution in the rounded solution is strictly larger than a
$\frac{3}{4}$ fraction of their contribution in the fractional
solution).
 
\subsection{Analysis for primary assignment}
 
First, let us prove a lemma which refers to the value recovered through primary assignment. Let
$x^p$ be the rounded solution before the secondary assignment step; then
 
\begin{lemma}\label{lem:primary-assignment}
For all $i\in\players$ and $j\in\items$, the expected contribution of $j$ to the objective value
due to primary assignment on $i$ is at least
$$
\EX[\Val_{ij}(x^p)] =
\rho(1-\rho(1-\rho))\Avg_{ij}(y)=(1-\rho(1-\rho))\Val_{ij}(y)
$$
where $\rho=x_{ij}$.
\end{lemma}

\begin{proof}
Let $i,i'$ be the players $j$ can be assigned to.  The probability
that $j$ belongs to the configuration picked by player $i$ is $x_{ij}
= \rho$.  If $j$ is picked by $i$,
the probability that $j$ is not picked by $i'$ is $\rho$, while the
probability that $j$ is also picked by $i'$ but is assigned to $i$ is
$(1-\rho)^2$. Therefore, the expected contribution of $j$ by the event
that $j$ is primarily assigned to $i$ is
$$
\EX[\Val_{ij}(x^p)]=\rho(1-\rho+\rho^2)\Avg_{ij}(y) =
(1-\rho+\rho^2)\Val_{ij}(y)
$$
since, conditioned on the event that $j$ is picked by $i$, the probability of picking a fixed
configuration $\conf\ni j$ is proportional to $y_{i\conf}$.
\end{proof}

The above lemma implies the following corollary:
\begin{corollary}\label{cor:primary-assignment}
Let $j\in\items$ can be assigned to players $i, i'$. The expected contribution of $j$ to the
objective value, if we only consider primary assignments, is
$$
\EX[\Val_j(x^p)] = (1-\rho(1-\rho))\Val_{ij}(y)+(1-\rho(1-\rho))\Val_{i'j}(y).
$$
where $\rho=x_{ij}$. Moreover, for any $\rho$,
$$
\EX[\Val_j(x^p)]\geq\frac{3}{4}\Val_{j}(y).
$$
\end{corollary}
 
\begin{proof}
The expected contribution of $j$ to the objective value of the
LP, considering only primary assignments, is
$$
\EX[\Val_{j}(x^p)]=\EX[\Val_{ij}(x^p)+\Val_{i'j}(x^p)]=(1-\rho+\rho^2)(\Val_{ij} (y) +
\Val_{i'j}(y))\geq \frac{3}{4}\Val_{j}(y).
$$
\end{proof}
 
Hence, we can deduce the following:
 
\begin{corollary}\label{cor:integral-items}
For all $j\in\items$, let $\rho_{\max}=\max_{i\in \players} \{x_{ij}\}$. Then, for all
$i\in\players$:
$$\EX [\Val_{ij}(x^p)]\geq (1-\rho_{\max}(1-\rho_{\max}))\Val_{ij}(y)$$
Furthermore
$$
\EX [\Val_{j}(x^p)]\geq (1-\rho_{\max}(1-\rho_{\max}))\Val_{j}(y).
$$
\end{corollary}
 
\subsection{Analysis for secondary assignment}
Now, let $\delta\in (0,\frac{1}{2})$ be a parameter to be defined later. Let
$I=\{j\in\items:\max\{x_{ij}:i\in \players\}\geq 1-\delta \}$ be the set of
almost-integral items and let
$H=\items\setminus I$ be the remaining items.
 
If all the items are almost-integral in our solution, it is
easy to design an algorithm which recovers more than a $\frac{3}{4}$-fraction of the value of each
one of them; this implies that when the fraction of the objective value which corresponds to
almost-integral items is non-negligible, we can immediately improve over the approximation
ratio $\frac{3}{4}$.Hence, our troubles begin when the contribution of almost-integral items to the
objective value is tiny.
 
Let $x^*$ be the rounded integral solution after the secondary assignment step. First, we prove a
lemma
concerning the value we gain from secondary assignment, and then conclude the analysis by proving
that the approximation ratio of our algorithm is greater than $\frac{3}{4}$.
 
\begin{lemma}\label{lem:half-integral-items}
For all $j\in H$, conditioned on the event that $j$ secondarily assigned to player $i$, who
initially picked configuration $\conf$,
$$\EX [\Val_j(x^*)]\geq \frac{\delta(1-\delta)\Val_{j}(y)}{2B_i}\sum_{j'\in \conf\cap
H}p_{ij'\conf}.$$
\end{lemma}
 
\begin{proof}
The main idea is the following: assume that all items belong to $H$,
and some item $j\in H$ fails to be primarily assigned (an event that
happens with constant probability). Then, we can take a look at player
$i$, to whom item $j$ is secondarily assigned; since we assumed all
the items belong to $H$, the items that belong to the configuration
$\conf$ that was picked by $i$ will be primarily assigned elsewhere
with constant probability, due to conflict resolution. Hence, there
will be on expectation a constant fraction of the budget of $i$ which
will be left free; this is the fraction of the budget that secondarily
assigned items will use to contribute to the objective value (see also
Figure~\ref{fig:nonunigraph}).

Let us now proceed with proving the lemma formally: let $H_i$ be the
items of $H$ that are secondarily assigned to player $i$, and hence do not
belong to $\conf$. By the definition of $x$, $\sum_{j'\in H_i}
\Val_{ij'}(y)\leq B_i$. For each $j'\in \conf\cap H$ and $j''\in H_i$,
we allocate a $\frac{\Val_{ij''}(y)}{B_i}$ fraction of the budget
which is left free when $j'$ is not assigned to $i$. Since every
$j'$ is not assigned to $i$ with probability $x_{ij'}(1-x_{ij'})\geq
\delta(1-\delta)$, and applying linearity of expectation, the expected
value of $j$ due to secondary assignment is at least
$$
\sum_{j'\in
C\cap H}p_{ij'\conf}x_{ij'}(1-x_{ij'})\frac{\Val_{ij}(y)}{B_i}\geq
\sum\limits_{j'\in
C\cap H}p_{ij'\conf}\delta(1-\delta)\frac{\Val_{ij}(y)}{B_i}
$$
 
Since
$\Val_{ij}(y)\geq\frac{\Val_{j}(y)}{2}$, the expected contribution of $j$ to the
objective value due to secondary assignment is at least
$$
\sum_{j'\in C\cap H}p_{ij'\conf}\delta(1-\delta)\frac{\Val_{j}(y)}{2B_i}
$$

\begin{figure}[!h]
\includegraphics[scale=0.7]{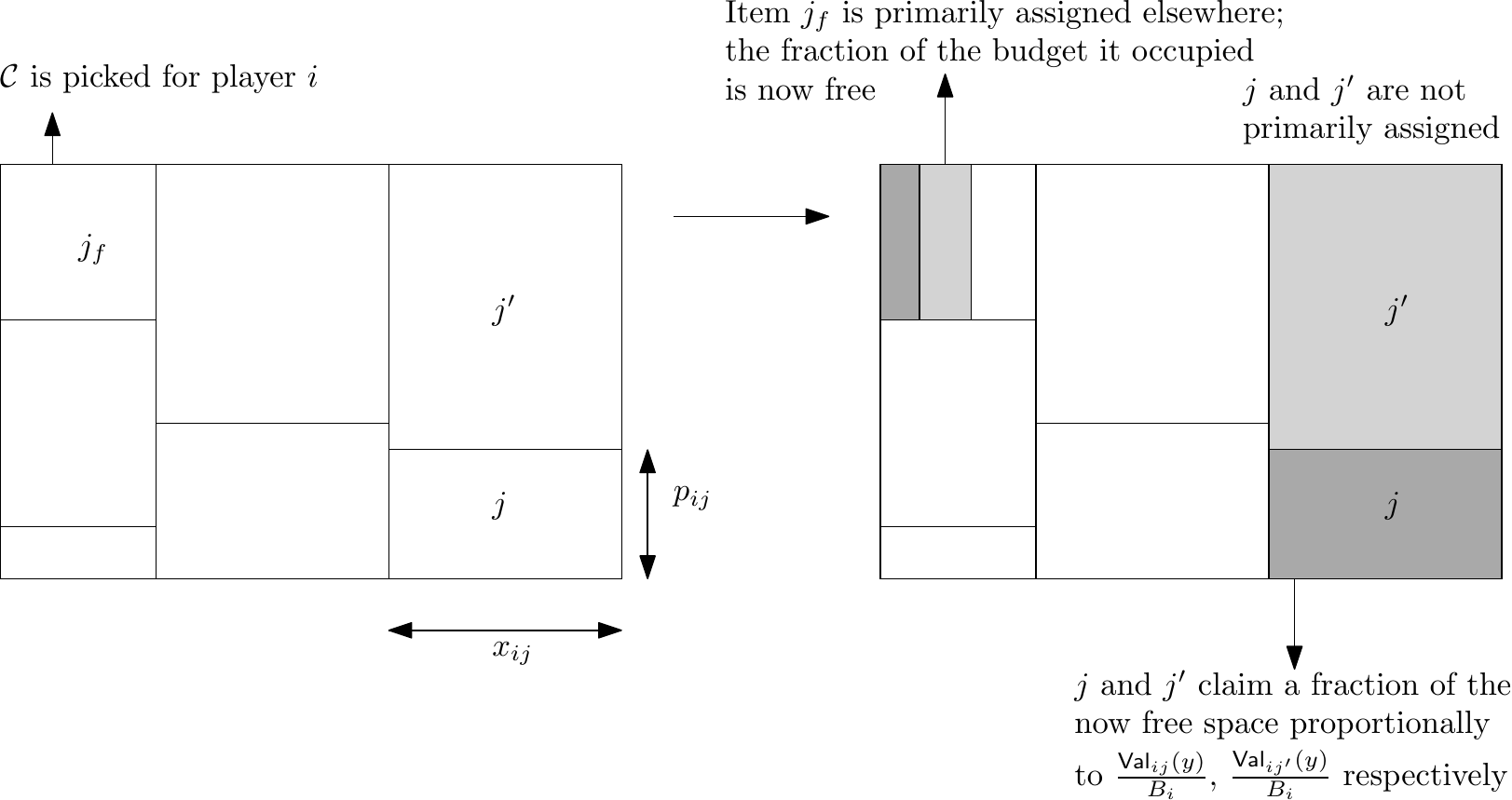}
\caption{Consider on the left the above fractional assignment of configurations to player $i$;
width
corresponds to the assignment value of an item and height to its price. Here, $\conf$ is picked for
player $i$. Assuming all the items belong to $H$, every item in $\conf$ has a constant probability
of being assigned elsewhere; in this case, it was $j_f$. Then, every item which does not belong to
$\conf$ is secondarily assigned to $i$ with constant probability(in this case $j,j'$) and claim
a constant fraction of the free space, thus increasing its contribution to the objective value.}
\label{fig:nonunigraph}
\end{figure}
 
\end{proof}
 
We are now ready to prove Theorem~\ref{thm:threshold}. By Corollary~\ref{cor:integral-items} we are
able to achieve a better than $\frac{3}{4}$ approximation guarantee when all items belong to $I$.
The fractional assignments of
items in $I$ are away from $\frac{1}{2}$ and therefore primary
assignment is sufficient to return better than $3/4$-fraction of their contribution to the objective
value. On the other hand, we get approximation guarantee greater than $\frac{3}{4}$ if all items
belong to $H$ due to Lemma~\ref{lem:half-integral-items}. The remaining scenario is when we have
items from both $H$ and $I$.
 
\begin{proof}[Proof of Theorem~\ref{thm:threshold}]
By definition,
$$\Val(x^*)=\sum_{j\in\items}\Val_{j}(x^*)=\sum_{j\in I}\Val_{j}(x^*)+\sum_{j\in H}\Val_{j}(x^*)$$
 
Now, remember that $j\in H$ is primarily assigned with probability $1-x_{ij}+x_{ij}^2$ and
secondarily assigned with probability $x_{ij}(1-x_{ij})\geq \delta(1-\delta)$, where $i$ is one of
the two players it can be assigned to. Let $s(j)$ be the player $j$ is secondarily assigned to when
needed, and let $\rho_j=x_{s(j)j}$. From Lemmas~\ref{lem:primary-assignment}
and \ref{lem:half-integral-items} we have
$$
\EX[\Val(x^*)]\geq \sum_{i\in
\players}\sum_{j\in I}(1-\delta+\delta^2)\Val_{ij}(y)+
$$
$$\sum_{j\in
H}(1-\rho_{j}(1-\rho_{j}))\Val_{j}(y)
+\frac{\rho_j(1-\rho_j)\Val_j(y)}{\sum_{\conf\subseteq\items:j\notin
\conf}y_{s(j)\conf}}\sum_{\conf\subseteq\items:j\notin
\conf}y_{s(j)\conf}\frac{\delta(1-\delta)}{2B_{s(j)}}\sum_{j'\in \conf\cap H}p_{s(j)j'\conf}=
$$
$$
\sum_{i\in \players}\sum_{j\in I}(1-\delta+\delta^2)\Val_{ij}(y)+
\sum_{j\in H}\frac{3}{4}\Val_{j}(y)
+\rho_j\Val_j(y)\sum_{\conf\subseteq\items:j\notin
\conf}y_{s(j)\conf}\frac{\delta(1-\delta)}{2B_{s(j)}}\sum_{j'\in \conf\cap H}p_{s(j)j'\conf}
$$
Let $\lambda=\delta(1-\delta)$; the sum corresponding to items in $H$ now becomes
$$
\sum_{j\in H}\frac{3}{4}\Val_{j}(y)
+\frac{\rho_j\Val_j(y)\lambda}{2B_{s(j)}}\sum_{\conf\subseteq\items:j\notin
\conf}y_{s(j)\conf}\left(\sum_{j'\in \conf}p_{s(j)j'\conf}-\sum_{j'\in \conf\cap
I}p_{s(j)j'\conf}\right)
$$
Our next goal will be to cancel out the negative term in the above sum; in order to do this, we
will manipulate the sum corresponding to items in $I$ as follows: for all $i\in\players$, $j\in I$
we will remove a
quantity equal to $\lambda\Val_{ij}(y)$ from the sum corresponding to items in $I$ and distribute it
to items
$j'\in H$ which are assigned to $i$ in configuration $\conf$ with a coefficient of
$p_{ij\conf}y_{i\conf}$. So, the sum over almost-integral items becomes
$$
\sum_{i\in \players}\sum_{j\in I}(1-\delta+\delta^2-\lambda)\Val_{ij}(y)
$$
which means that choosing the right $\delta$ we will still be able to recover more than
$\frac{3}{4}$ of the value of every item in
$I$. On the other hand, because of this redistribution, we can consider that for any player $i$
the total subtracted value from items in $I$ is distributed to items assigned in $i$ proportionally
to their fractional assignment; hence, we can actually consider that this quantity is distributed
among configurations proportionally to their assignment and afterwards distributed among items in
that configuration. In other words, for every item $j\in H$ and every configuration $\conf$
that $j$ belongs to in $s(j)$, and every $j'\in\conf\cap I$, we will remove a $\lambda$
fraction of $y_{s(j)\conf}p_{s(j)j'\conf}$(i.e. the contribution of $j'$ to the objective value due
to $\conf$) and assign a $\frac{\Val_{s(j)j}(y)}{B_{s(j)}}$ fraction of it towards $j$. Hence, since
$\Val_{s(j)j}(y)\geq \frac{\Val_{j}(y)}{2}$, the sum over items in $H$ becomes
$$
\sum_{j\in H}\frac{3}{4}\Val_{j}(y)
+\frac{\rho_j\Val_j(y)\lambda}{2B_{s(j)}}\sum_{\conf\subseteq\items:j\notin
\conf}y_{s(j)\conf}\sum_{j'\in \conf}p_{s(j)j'\conf}
$$
 
 
Since if there is a configuration which does not contain items of total value at least the budget
of the corresponding player, our algorithm can only perform better (there is more leftover
space than that which we have estimated), we can assume without loss of generality that for all
$\conf\subseteq \items$ for which there is a player $i$ such that $y_{i\conf}>0$, it holds
$\sum_{j\in \conf}p_{ij\conf}= B_i$. Hence, the sum over items in $H$ becomes
$$
\sum_{j\in H}\frac{3}{4}\Val_{j}(y)
+\frac{\rho_j\Val_j(y)\lambda}{2}\sum_{\conf\subseteq\items:j\notin
\conf}y_{s(j)\conf}=
\sum_{j\in H}\frac{3}{4}\Val_{j}(y)
+\frac{\rho_j(1-\rho_j)\Val_j(y)\lambda}{2}\geq
$$
$$
\sum_{j\in H}\frac{3}{4}\Val_{j}(y)
+\frac{\delta(1-\delta)\Val_j(y)\lambda}{2}
$$
 
In total we have:
$$
\EX[\Val(x^*)]\geq
\sum_{i\in \players}\sum_{j\in I}(1-\delta+\delta^2-\lambda)\Val_{ij}(y)+
\sum_{j\in H}\frac{3}{4}\Val_{j}(y)
+\frac{\lambda\delta(1-\delta)}{2}\Val_j(y)=
$$
$$
\sum_{j\in I}(1-\delta+\delta^2-\lambda)\Val_{j}(y)+
\sum_{j\in H}\frac{3}{4}\Val_{j}(y)
+\lambda\delta(1-\delta)\Val_j(y)
$$
 
Selecting $\delta$ such that $(1-\delta+\delta^2-\lambda) =(1-\delta+\delta^2-\delta(1 - \delta))
>\frac{3}{4}$, the theorem follows.
\end{proof}

%% file: gap.tex
\section{An improved integrality gap for the unrestricted case}
\label{sec:gap}

The previously best known upper bound on integrality gap of the configuration LP was $5/6 = 0.833$
proved by Chakrabarty and Goel in \cite{DBLP:journals/siamcomp/ChakrabartyG10}. We improve this to
$0.828$. Unlike the previous result, our gap instance is not a graph instance.

\begin{theorem}
  The integrality gap of the configuration LP is at most $2(\sqrt{2} - 1) \approx 0.828$.
\end{theorem}

\begin{proof}
For $p,q\in\mathbb{N}$ such that $p<q$, consider the following budget assignment instance: there
are $q$ players $b_i\in B$ for $1\leq i\leq q$ with budget 1 and $q$ players $s_i\in S$ for $1\leq
i\leq q$ with budget $\frac{p}{q}$. Additionally, there are $p$ items $c_j\in C$ for $1\leq j\leq
p$, each of which can be assigned to players in $B$ with a value of 1. Finally, for each player
$b_i\in B$, there are $q$ items $o_{i,j}\in O_i$ for $1\leq j\leq q$, which can be assigned to $b_i$
and $s_i$ with a value of $\frac{1}{q}$.

An example with $p = 2$ and $q=3$ is drawn in Figure~\ref{fig:gap}.

\begin{figure}[h]
  \begin{center}
    \includegraphics{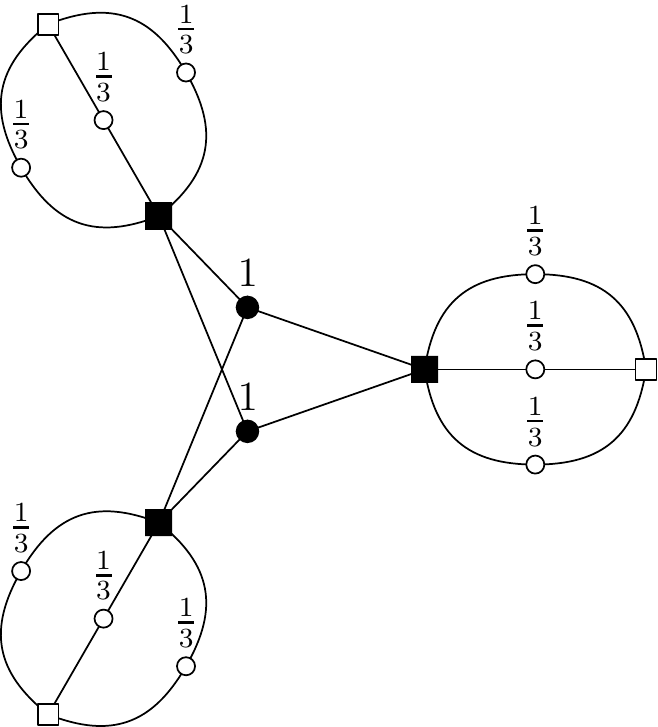}
  \end{center}
  \caption{An instance with $p = 2$ and $q = 3$. Black squares are players in $B$, white squares
    players in $S$. They have budgets $1$ and $2/3$ respectively. Items in $C$ are black dots and
    items in $O_i$ are white dots with their values written next to them. An edge between a player
    and an item denotes that the player is interested in that item.}
  \label{fig:gap}
\end{figure}

The optimal integral solution assigns $p$ items from $C$ to $p$ distinct players in $B$; for each
player $i$ that is assigned an item from $C$, we assign $p$ items
from $O_i$ to $s_i$. Let $i'$ be one of the $q-p$ players which do not get an item from $C$, the
optimal integral solution assigns the $q$ items from $\bigcup_{i=1}^{q} O_i$ to $i'$. The total value
of the solution is $p(1+\frac{p}{q})+q-p$.

Consider the following fractional solution to the configuration LP. Every item in $C$ is shared by
the $q$ players in $B$, each with a fraction of $\frac{1}{q}$. Furthermore, every player $i$ in
$B$ is assigned a fraction $\frac{q-p}{q}$ of every item in $O_i$. More precisely, the configuration
$\conf = O_i$ has $y_{i\conf} = \frac{q-p}{q}$, so the budget of $i$ is completely filled.

Finally, every player $i$ in $S$ uses the unassigned fraction $\frac{p}{q}$ of every item in $O_i$
to form $p\choose q$ configurations of size $\frac{p}{q}$, which fill up the budget of $i$
completely. Hence, the value of the fractional solution is $q+q\frac{p}{q}=q+p$. Note that the total
value of items is $p+q$, so there can not be a better assignment.

Hence, the integrality gap $I(p, q)$ is
\begin{equation*}
I(p,q)=\frac{p(1+\frac{p}{q})+q-p}{p+q}=\frac{p^2+q^2}{q^2+pq}=\frac{\frac{p^2}{q^2}+1}{\frac{p}{q}+
1}.
\end{equation*}

For $x\in\mathbb{R}$, the expression $f(x)=\frac{x^2+1}{x+1}$ is minimized at $x^*=\sqrt{2}-1$ and
has $f(x^*)=2(\sqrt{2}-1)$. Hence, choosing $p,q$ such that $\frac{p}{q}$ is arbitrarily close to
$\sqrt{2}-1$, we can achieve an integrality gap arbitrarily close to $2(\sqrt{2}-1)$.
\end{proof}

%% file: hardness.tex
\newcommand{\mlinthree}{\textsc{Max-3-Lin(2)}}
\newcommand{\mlintwo}{\textsc{Max-2-Lin(2)}}

\section{Hardness of Approximation}
\label{sec:hardness}

In this section we strengthen the known hardness results. First we prove that the known
$15/16$-hardness holds also for restricted MBA where players have the same budget and then we prove
hardness $59/60$ in the graph case.

\begin{theorem}
  For every $\epsilon > 0$, it is NP-hard to approximate restricted MBA within a factor of $15/16 + \epsilon$.
  Furthermore, this is true for instances where all items can be assigned to at most 3 players and
  all players have the same budget.
\end{theorem}
\begin{proof}
  Chakrabarty and Goel in \cite{DBLP:journals/siamcomp/ChakrabartyG10} prove the $(15/16 +
  \epsilon)$-hardness for restricted MBA instances where all items can be assigned to at most 3
  players. They achieve this by reducing \mlinthree problem to MBA. The \mlinthree problem was
  proved to be NP-hard to approximate within a factor of $1/2+\epsilon$ by H\aa stad in
  \cite{DBLP:journals/jacm/Hastad01}.

  We use the same proof but use a different starting point. The result of H\aa stad can be modified
  with the technique of Trevisan~\cite{DBLP:conf/stoc/Trevisan01} so that each variable
  in the \mlinthree instance has the same degree.

  The construction of Chakrabarty and Goel gives budget $4 \deg (x)$ to the 2 players
  corresponding to variable $x$. Hence, if all variables have the same degree, all players have the
  same budget.
\end{proof}

Next, we modify the construction of Chakrabarty and Goel for use with linear equations of size 2.
The important change is that we create items for assignments that do not satisfy an equation, while
previous construction used satisfying assignments. The use of equations of size $2$ implies a
hardness for the graph case, i.e.\ where each item can only be assigned to two players.

\begin{theorem}
  For every $\epsilon > 0$, it is NP-hard to approximate graph MBA within a factor of $59/60 +
  \epsilon$. Furthermore, this is true for the restricted instances where all players have the same
  budget.
\end{theorem}

\begin{proof}
  We reduce from an instance $\phi$ of \mlintwo. Let $x$ be a variable occurring $\deg(x)$ times in
  $\phi$. We have two players $\left<x : 1\right>$ and $\left<x : 0\right>$ both with budgets
  $\deg(x)$ and an item of value $\deg(x)$ that can only be assigned to these two players. The meaning
  of this item is that if it is assigned to the player $\left<x: a\right>$, then a truth assignment
  $\alpha$ has $\alpha(x) = a$.

  For each equation $x + y = b$, there are two items $\left<x: a_1, y : a_2\right>$ of value $1$.
  Each such item corresponds to an assignment $\alpha$ for which $\alpha(x) = a_1,
  \alpha(y) = a_2$ and $a_1 + a_2 \not = b$. An item $\left<x: a_1, y : a_2\right>$ can be assigned
  to $\left<x : a\right>$ and $\left<y : c\right>$ only if $a = a_1$ and $c = a_2$ respectively.

  Every item can only be assigned to two players, so this is a graph instance. Furthermore, the
  valuation for both players is the same, so it is the restricted case.

  The analysis is now very similar to the one in \cite{DBLP:journals/siamcomp/ChakrabartyG10}. We
  can prove that an optimal assignment of items always assigns items of weight $\deg(x)$ and this
  can be translated into a truth assignment $\alpha$ to variables. We have $\alpha(x) = a$ if an
  item of value $\deg(x)$ is assigned to $\left<x : a\right>$. If $\alpha$ satisfies $x + y = b$,
  i.e.\ $\alpha(x) + \alpha(y) = b$, we can assign both items $\left<x: a_1, y : a_2\right>$.
  Otherwise we can only assign one of them, since both $\left<x: \alpha(x) \right>$ and $\left<y:
  \alpha(y)\right>$ are fully assigned. So if $\phi$ is $\delta$-satisfiable with $m$ equations, the
  MBA instance has objective value $\sum_x \deg(x) + m(2\delta + (1-\delta)) = 3m
  + \delta m$.

  H\aa stad and Trevisan et al.\ in \cite{DBLP:journals/jacm/Hastad01} and
  \cite{DBLP:journals/siamcomp/TrevisanSSW00} proved that it is NP-hard to distinguish instances of
  \mlintwo that are at least $(3/4 - \epsilon)$-satisfiable and those that are at most $(11/16 +
  \epsilon)$-satisfiable. Hence it is hard to distinguish between an instance of MBA with objective
  value at least $3m + \frac{3}{4}m - \epsilon m=m (60/16-\epsilon)$ and at most $3m + \frac{11}{16}
  m + \epsilon m = m (59/16+\epsilon)$, where $m$ is the number of equations in $\phi$.

  Therefore graph MBA is NP-hard to approximate to within a factor of $59/60 + \epsilon$. The
  instance from \cite{DBLP:journals/siamcomp/TrevisanSSW00} can also be modified to be regular,
  i.e.\ with all degrees $\deg(x)$ the same, thus producing an instance with equal budgets. This can
  be done by splitting each variable into more variables, as in \cite{DBLP:conf/stoc/Trevisan01}.
\end{proof}

%% file: future.tex
\section{Conclusion and future directions}

We showed that the integrality gap for configuration LP is strictly better than $\frac{3}{4}$ for
two interesting and natural restrictions of Maximum Budgeted Allocation: restricted and graph MBA.

These results imply that the configuration LP is strictly better than the natural assignment LP and
pose promising research directions. Specifically, our results on restricted MBA suggest that our limitations in
rounding configuration LP solutions do not necessarily stem from the items being fractionally
assigned to many players, while our results on graph MBA suggest that they do not necessarily stem from the items
having non-uniform prices. Whether these limitations can simultaneously be overcome is left as an interesting open problem.

Finally, it would be interesting to see whether the techniques presented, and
especially the exploitation of the big items structure, can be applied to other allocation problems
with similar structural features as MBA (e.g. GAP).